\documentclass[10pt,journal,compsoc]{IEEEtran}
\ifCLASSOPTIONcompsoc
  \usepackage[nocompress]{cite}
\else
  \usepackage{cite}
\fi

\ifCLASSINFOpdf
 
\else
 
\fi

\usepackage[small]{caption}
\usepackage{amsmath}
\usepackage{mathtools}
\usepackage{algorithm}
\usepackage{graphics,graphicx,dblfloatfix}
\usepackage{amsthm}
\RequirePackage[table]{xcolor}
 \usepackage{textcomp}
 \usepackage{algpseudocode}
 \usepackage{ragged2e}

 \newlength\myindent
\newtheoremstyle{component}{}{}{}{}{\bfseries}{.}{.5em}{\thmnote{#3}#1}
    \theoremstyle{component}

\begin{document}

\title{A Novel Information Theoretic Framework for Finding Semantic Similarity in WordNet }

\author{Abhijit~Adhikari,
        Shivang~Singh,
        Deepjyoti~Mondal,
        Biswanath~Dutta,
	Animesh~Dutta
\IEEEcompsocitemizethanks{\IEEEcompsocthanksitem Abhijit~Adhikari is with the Department
of Information Technology, National Institute of Technology, Durgapur, West Bengal, India 713209.\protect\\

E-mail: abhijitbitmesra@gmail.com
\IEEEcompsocthanksitem Shivang~Singh is with Amdocs, India.\protect\\
E-mail:shivangsingh777@gmail.com
\IEEEcompsocthanksitem Deepjyoti~Mondal is with the Department
of Information Technology, National Institute of Technology, Durgapur, West Bengal, India 713209.\protect\\
E-mail:djmdeveloper060796@gmail.com
\IEEEcompsocthanksitem Biswanath~Dutta is with the DRTC, Indian Statistical Institute, Bangalore, Karnataka, India 560059.\protect\\
E-mail:dutta2005@gmail.com
\IEEEcompsocthanksitem Animesh~Dutta is with the Department
of Information Technology, National Institute of Technology, Durgapur, West Bengal, India 713209.\protect\\
E-mail:animeshnit@gmail.com
}
}

\IEEEtitleabstractindextext{
\begin{abstract}\emph{Information content} (IC) based measures for finding \emph{semantic similarity} is gaining preferences day by day. 
Semantics of concepts can be highly characterized by \emph{information theory}. The conventional way for calculating IC is based on the probability 
of appearance of concepts in corpora. Due to data sparseness and corpora dependency issues of those conventional approaches, 
a new corpora independent \emph{intrinsic} IC calculation measure has evolved. In this paper, we mainly focus on such intrinsic IC model and several topological
aspects of the underlying ontology. Accuracy of intrinsic IC calculation and semantic similarity measure rely on these aspects deeply.
Based on these analysis we propose an information theoretic framework which comprises an intrinsic IC calculator and a semantic similarity model.
Our approach is compared with state of the art 
semantic similarity measures based on corpora dependent IC calculation as well as intrinsic IC based methods using several benchmark data set. 
We also compare our model with the related Edge based, Feature based and Distributional approaches. 
Experimental results show that our intrinsic IC model gives high correlation value when applied to different semantic similarity models.
Our proposed semantic similarity model also achieves significant results when embedded with some state of the art IC models including our's.
\end{abstract}
}

\maketitle

\IEEEdisplaynontitleabstractindextext

\IEEEpeerreviewmaketitle

\IEEEraisesectionheading{\section{Introduction}\label{sec:introduction}}

\IEEEPARstart{S}{emantic measures}\cite{38} are widely accepted nowadays to evaluate the proximity of semantic relationship between 
elements of various types: units of language, diseases, genes, geographical locations and so on. 
There are two main notions in semantic measures. One is semantic similarity \cite{38}, \cite{47} and another is
\emph{semantic relatedness}\cite{38}, \cite{47}. Actually in informal way both describes how concept A is related to concept B. 
But there is a significant difference between semantic similarity
and semantic relatedness. Semantic similarity considers only taxonomical relationships for measuring the semantic strength between two 
concepts\cite{47}, e.g. rafting and water polo both are similar because both are water sports. Whereas semantic relatedness considers taxonomic and 
non taxonomic relations (e.g. meronymy\cite{47}, functionality, cause effect, etc.) between concepts\cite{47}, e.g. food poison and stomach pain both are related.
Food poison is the cause of stomach pain. In this paper, our concern is semantic similarity between two concepts.

In a nutshell semantic similarity is more difficult to model than semantic relatedness, because later 
is a more general in relationship \cite{41}.
Semantic similarity between concepts has high importance from many years in artificial 
intelligence and cognitive science. It has been successfully applied directly or indirectly in several areas
like word sense disambiguation \cite{23}, synonym detection \cite{3}, automatic spelling error detection 
and correction \cite{24}, thesauri generation \cite{25}, information extraction \cite{26, 27}, semantic annotation \cite{28}, 
ontology merging \cite{29} etc.. Semantic similarity has been also applied in the field of formal concept analysis \cite{30} like, 
clustering of structured resources \cite{31}, question answering \cite{32}, 
development of recommender systems \cite{33}, and multi-agent systems \cite{34}. In the field of geo-informatics 
semantic similarity is used to compute how well two geographic concepts are related based on their domain specific ontology
like geo-net-pt \cite{37}. There are several biomedical ontologies available nowadays. 
Some well known examples are SNOMED-CT \cite{35}, MeSH \cite{36} etc. Due to availability 
of such large well structured resources in biomedical domain, interests have been growing in finding semantic similarity assessments 
based on these ontologies in biomedical domains. They are mainly used to compare genes and proteins based on similarity of their
functions rather than on their sequence similarity \cite{54}.

Several measures for finding semantic similarity between two concepts have been proposed till now. 
Those measures are broadly classified as \emph{distributional measure} \cite{38} and \emph{knowledge based measure} \cite{38}.
Distributional measure totally relies on corpus and has several limitations. 
Indexing and ranking mechanisms used in search engines make distributional measure biased in nature \cite{41}. 
Statistical distribution based on words of this measure also overlooks the fact 
that the semantic units can be Multi Word Entity (MWE)\cite{41}.  
Beside these, in distributional measure the words to be compared must have to occur at 
least few times in the corpus.
Even sense tagged corpora are hardly available for comparing semantics between two concepts by this measure \cite{38}. 
Due to availability of large number of ontologies in several domains, semantic similarity based on knowledge based measures
are gaining preferences over distributional measures. Knowledge based measure depends on user defined resources such as thesauri, 
taxonomies or encyclopedias, as the context of comparison \cite{41}.

Several approaches in knowledge based paradigm are available 
like, \emph{edge-based} \cite{38} approach considers the similarity as a function of distance which separates two concepts in the ontology and
\emph{feature-based} \cite{38} strategy evaluates a concept as a set of features. The features of a concept are usually 
considered as the set of concepts subsuming it, i.e. its ancestors. Another approach from knowledge based measures relies 
on information theory \cite{38} which assesses the similarity of concepts according
to the amount of information they provide, i.e. their information content (IC). This IC measure is one of the best measures
among all proposed ones. The similarity value gained by this measure is more accurate and effective than others.
This information theoretic approach for
finding semantic similarity is actually a two folded process. First step is to calculate IC of each concepts. In second step these IC values are used to
calculate semantic similarity between any two concepts. In initial stages, calculation of IC relies completely on corpus,
i.e. IC of any concept is calculated based on the frequencies of that concept evaluated from the corpus where the concept or its 
any of the instance is used. Problem of this method is, it suffers from data sparsity. 
Beside this, for having tagged corpora we need huge human efforts though, accuracy of this measure is higher 
than distributional measures. To overcome the loopholes of the earlier
IC calculation technique, an intrinsic IC calculation technique has been evolved. This intrinsic approach totally relies on 
the ontology that is used for finding semantic similarity. This approach out-performs the existing corpus based IC calculation 
techniques.
An efficient semantic similarity measure embedded with an efficient intrinsic IC calculator as a whole can out perform 
other state of the art semantic similarity measures. 

Our main contributions of this paper are given below:
\begin{itemize}
 \item This paper is an extension work of our earlier research \cite{40}. 
In that work, we have proposed a novel intrinsic IC calculator. In the earlier research we have tested our IC calculator by three classical
semantic similarity calculators based on a smaller benchmark data set. In our current paper we check our IC model more thoroughly using bigger benchmark 
data set. We also evaluate performance by embedding proposed IC model with some more state of the art semantic similarity checkers. 
It is shown in this paper that our IC model with some specific semantic similarity calculator gives better results than
any of the state of the art IC calculation models.
\item This paper also proposes a novel technique for measuring semantic similarity between two concepts. Our semantic similarity calculator gives
significant similarity scores based on the proposed IC calculator. Others IC calculator also gives significant results when applied to our semantic similarity
calculator.
\end{itemize}

\indent The rest of the paper is organized as follows: In section 2 we discuss about previous works
in related domain. Section 3 defines scope of the current work. Section 4 discusses the proposed system model and metric used. Section 5 and 6 
describe the proposed solution, and experiments respectively. Section 7 concludes the paper.

\section{Related Work}
Finding semantic similarity based on IC is basically a two folded process. Those are calculating IC and
measuring semantic similarity based on calculated IC.

\textit{A. IC calculation model:}
The core part of calculating semantic similarity of two concepts depends on finding IC first. 
The more accurate the IC calculation technique is, the more 
accurate evaluated similarity value would be. So, calculating IC with 
more perfection is very much crucial. This IC calculation technique also can be divided as

\textit{A.1 Corpora based IC calculation techniques:}

Resnik \cite{1} has first proposed corpora based IC calculation measure. This type of \emph{information theoretic}
approaches for calculating IC of any concept, depend on the inverse of the 
probability of that concept's frequency in the underlying corpus where it is used: 
\begin{equation}
\label{eqn:Corporabasedic}
  IC(c) = -\log(p(c)) 
\end{equation}
So, a concept 
which is more frequent has less IC than a concept having less 
frequency. IC of a concept from an ontology monotonically decreases as one goes 
towards root node. For calculating IC of a concept of any ontology depends on all of its 
taxonomical hyponyms' frequencies in the corpus. So, probability of that concept is calculated in the following way:
\begin{equation}
\label{eqn:probability}
 p(c)= \frac {\Sigma_{w\in W(c)} freq(w)}{N}
\end{equation}
where, $W(c)$ is the set of terms in the corpus whose senses are subsumed by concept $c$.
$N$ is the total number of corpus terms contained in the taxonomy.
Lin \cite{3}, Jiang and Conrath \cite{2} have extended Resnik's similarity measures but have used Resnik's Corpora based IC calculation techniques.
For the accurate computation of concept appearance probabilities, 
word sense must be determined in the corpus as textual corpora contains words. These concepts are modeled by ontology. 
This process needs to remove ambiguation and requires proper annotation of each concept found in the corpus. 
Any kind of changes made in the taxonomy or in the corpus, recomputation 
is needed each time for the affected concepts. This process is time consuming and need human intervention.
The size and nature of the input highly effect the resulting probability. 
The ontology should be more complete. Corpora's
contents should be sufficient with respect to the ontology scope and
large enough to avoid data sparseness. Though Brown corpus \cite{22} may
be suitable for WordNet \cite{21} \cite{42} ontology, but more specific corpora is be
needed for domain ontologies. So, scalability issue and data sparseness \cite{5} hamper the applicability of this earlier IC calculation model.

\textit{A.2 Intrinsic IC calculation techniques:}

To overcome the loopholes mentioned in sub-section A.1, several authors \cite{1, 4, 5, 6, 7, 8} have proposed intrinsic IC computation model. In this model,
we do not have to rely on any corpora for calculating IC. IC is calculated based on the ontology itself. Seco et al. \cite{5} have formulated IC 
calculations technique intrinsically first. Seco's model relies on number of concept hyponyms of the underlying ontology in the following way:
\begin{equation}
\label{eqn:Secoic}
 IC_{seco}(c_i) = \frac{\log(\frac{hypo(c_i) + 1}{max_{wn}})}{\log(\frac{1}{max_{wn}})} 
\end{equation}
where, $hypo(c_i)$ represents number of hyponyms the concept $c_i$ has and $max_{wn}$ represents maximum number of concepts the taxonomy has.
WordNet is considered as underlying taxonomy in their approach. The problem of this model is, concepts of same number of hyponyms but 
different generality will result same similarity value. To tackle this problem Zhou et al. \cite{7} have introduced relative depth with number of 
hyponyms of the concept:
\begin{equation}
\label{eqn:Zhouic}
\footnotesize
\noindent IC_{zhou}(c_{i})=K(1\textendash \frac{\log(hypo(c_i)+1)}{\log(node_{max})}) + (1\textendash K)\frac{\log(deep(c_i))}{\log(deep_{max})}
\end{equation}
Introducing depth as weighted feature is problematic as the weight represents a parameter that must be empirically tuned. To avoid this problem 
S\'{a}nchez et al. \cite{6} have introduced a new intrinsic model for IC calculation:
\begin{equation}
\label{eqn:Davidic}
 IC_{david}(c_i)=\textendash \log(\frac{\frac{leaves(c_i)}{subsumers(c_i)}+1}{max_{leaves}+1})
\end{equation}
Still some problems remain in this model. Concepts having same number of subsumers and leaves but different hyponym structure and different
number of hyponyms will generate same IC values. That means both concepts have same meaning. But they should convey different information.
Meng et al. \cite{8} have proposed a new model again to overcome this above mentioned problem of \emph{eq.~\ref{eqn:Davidic}}. They have considered depth of the concept and depth of each hyponym
of the concept.
David S\'{a}nchez \cite{4} has proposed again a new IC calculation model based on commonness of a concept. Actually this model relies on number of subsumers 
of leaf nodes of the concept whose IC is being calculated in the following way:
\begin{equation}
\label{eqn:Davidcommonnessic}
 IC_{david}(c_i)=\textendash \log(\frac{commonness(c_i)}{commonness(root)})
\end{equation}
where, $commonness(c_i)=\Sigma commonness(l)$, $\forall$ $l$ $\mid$ $l$ is a leaf node $\Lambda$ $l$ is subsumed by concept $c_i$ and $c_i$ is not a leaf node.
$commonness(l)=\frac{1}{subsumers(l)}$.
Qingbo et al. \cite{50} have proposed a new intrinsic IC calculation model with some different topological factors in the following way:
\begin{equation}
\label{eqn:Qingboic}
 IC_{qingbo}(c_i)=f_{depth}(c_i)\ast(1-f_{leaves}(c_i))+f_{hypernyms}(c_i)
\end{equation}
where, $f_{depth}(c_i)$, $f_{leaves}(c_i)$, $f_{hypernyms}(c_i)$ are defined as follows
\begin{equation}
f_{depth}(c_i)=\frac{\log(depth(c_i))}{\log(max\_depth)}
\end{equation}

\begin{equation}
f_{leaves}(c_i)=\frac{\log(leaves(c_i)+1)}{\log(max\_leaves+1)}
\end{equation}

\begin{equation}
f_{hypernyms}(c_i)=\frac{\log(hyper(c_i)+1)}{\log(max\_nodes)}
\end{equation}
where $c_i$ is the evaluated concept in the ontology,
and $depth(c_i)$, $hyper(c_i)$ and $leaves(c_i)$
corresponds to its depth, the number of its hypernyms, and the number of its leaves
respectively. In addition, $max\_depth$, $max\_nodes$ and
$max\_leaves$ are three constants concerned with the
background ontology.

\textit{B. Semantic similarity measure based on IC:}
There are basically three classical semantic similarity measures available which are used by authors to judge the performance of
their IC models. Resnik \cite{1} has first proposed IC based semantic similarity measure. According to Resnik similarity
between two concepts depends upon IC of their \emph{least common subsumer} (LCS) (i.e. the concept which 
subsume both the concepts and has maximum depth among all other subsumers of those concepts) in the following way: 
\begin{equation}
\label{eqn:Ressim}
  Sim_{res}(c_i, c_j) = \max( IC(LCS(c_i, c_j))) 
\end{equation}
$Sim(c_i, c_j)$ represents semantic similarity between concepts $c_i$ and $c_j$ throughout this paper.
If the concepts under consideration do not have any common subsumers then their similarity will be considered as zero.
The problem of Resnik is any pair of concepts having same LCS 
will have same similarity value. To overcome this problem Lin \cite{3} and Jiang and
Conrath \cite{2} has extended Resnik's work by considering IC of the each evaluated concepts.
Lin considered Resnik's similarity formula and has made a 
ratio with summation of individual IC of each concept:
\begin{equation}
\label{eqn:Linsim}
  Sim_{lin}(c_i, c_j) = \frac{2 \times Sim_{res}(c_i, c_j)}{IC(c_i)+IC(c_j)}
\end{equation}
Jiang and Conrath have proposed similarity between two concepts in terms of distance between two concepts using IC:
\begin{equation}
\label{eqn:Jiangsim}
  Dist_{j\&c}(c_i, c_j) = IC(c_i) + IC(c_j) - 2 \times Sim_{res}(c_i, c_j) 
\end{equation}
If we take opposite of distance we will get similarity scores from Jiang and Conrath measure.

Apart from these three classical semantic similarity calculators Pirr\'{o} et al. \cite{48} and Montserrat Batet \cite{47} also have proposed some IC basd 
semantic similarity calculators. Pirr\'{o} et al. have presented a framework, which maps the feature-based model
of semantic similarity into the information theoretic domain. They call their model as FaITH model. The model looks like:
\begin{equation}
\label{eqn:Pirrosim}
  Sim_{FaITH}(c_i, c_j) = \frac{IC(LCS(c_i,c_j))} {IC(c_i) + IC(c_j) - IC(LCS(c_i,c_j))}
\end{equation}
Pirr\'{o} et al. have used IC model defined by Seco et al. in their FaITH model of semantic similarity.
Montserrat Batet \cite{47} mapped an edge counting semantic similarity measure into an IC based semantic similarity model in the following way:
\begin{equation}
\label{eqn:Batetsim}
  Sim_{m}=- \log \frac{IC(c_i) + IC(c_j) - 2\times IC(LCS(c_i,c_j))+1}{2\times max\_IC}
\end{equation}
In this model, Montserrat Batet has used S\'{a}nchez's IC model (\emph{equation \ref{eqn:Davidic}}) for calculating IC.
Apart from all these IC based semantic similarity models, many authors have proposed several non-IC based measures for calculating semantic similarity.
Rada et al. \cite{10}, Wu and Palmer \cite{11}, Leacock et al. \cite{12}, Li et al. \cite{13} are few of them who proposed edge-counting semantic similarity measures.
Rodriguez et al. \cite{14}, Tversky \cite{15}, and Petrakis et al. \cite{16} have proposed feature-based models for semantic similarity calculation. Bollegala et al.\cite{17}
Chen et al. \cite{18}, Sahami et al. \cite{19}, Gledson et al.\cite{45} are some of the authors who have proposed distributional semantic similarity
calculation models. Bollegala et al.\cite{44} has proposed again a Web Snippest and Page-count based approach where they have introduced some
clustering with their previous work\cite{17}. We do not discuss much about these non IC based semantic similarity model, as our main concern
is semantic similarity model based on intrinsic IC calculation technique.

\section{Scope of the Work}
According to the IC-model proposed by Meng et al., there is a possibility that
two concepts have same hyponym structure and stay in the same depth but with
different number of subsumers.
\begin{figure}[!htb]
 \begin{center}
 \includegraphics[scale=.60]{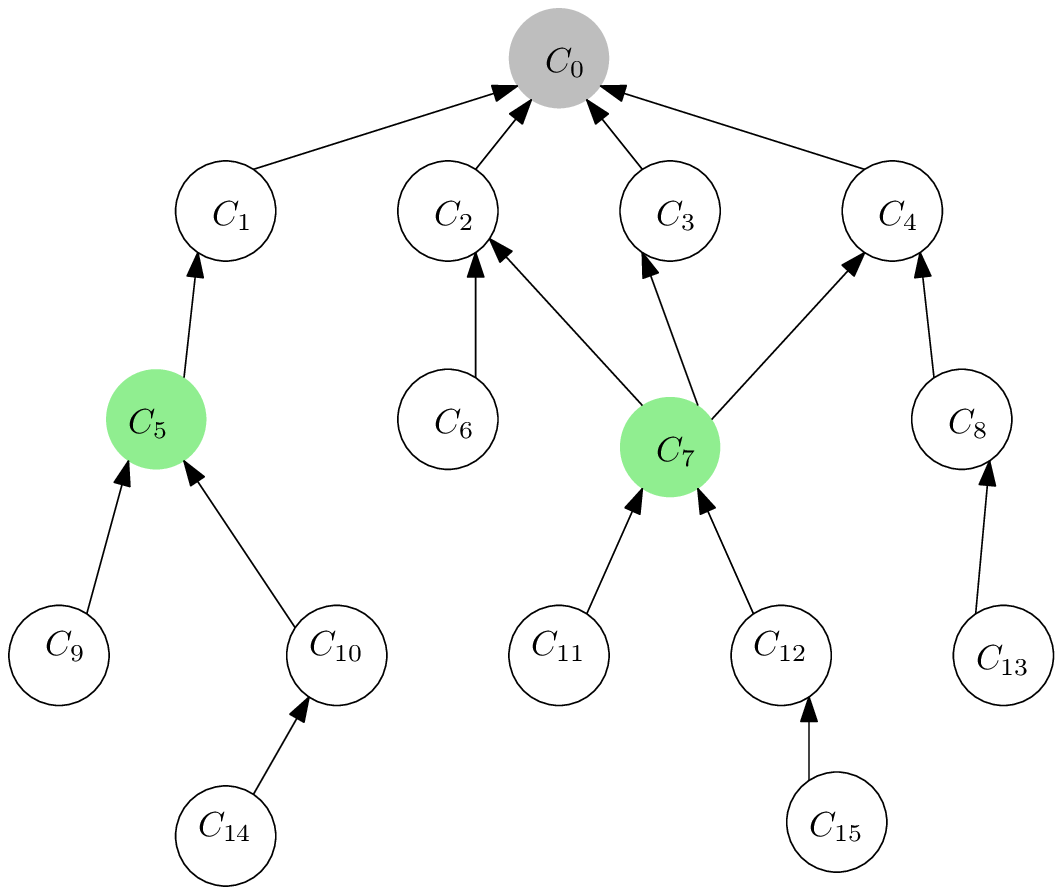}
 \caption{An example ontology of hyponymy-hypernymy relationship.}
 \end{center}
\end{figure}
\noindent 
In that case both concepts have also same IC value, which is not expected as the topological
structure of both concepts in the ontology is different.  
Consider $Fig.~1$, where concept $c_5$ and $c_7$ both have 
same depth. Both the concepts have same number of hyponyms and their structures are also same. 
But the differences present in the structure of subsumers which 
actually define the concreteness of any concept. Both have different number of subsumers. Topological arrangement of those subsumers are also different.
It is quiet obvious that the more we consider topological details, the more there is chances to get accurate IC values of concepts.
Actually in this model of Meng et al. both the concepts have different type of
inheritance. $c_5$ has no multiple inheritance and $c_7$ has multiple inheritances.
So, they should not result in same IC value because number of multiple inheritance 
of the concept should have some significant weightage in deciding IC intrinsically.
Again according to the model proposed by David S\'{a}nchez (\emph{eq.~\ref{eqn:Davidcommonnessic}}) \cite{4}, relies on only the number of subsumers 
of leaves of the concept whose IC is going to be calculated. But according	
\begin{figure}[!htb]
 \begin{center}
\includegraphics[scale=.60]{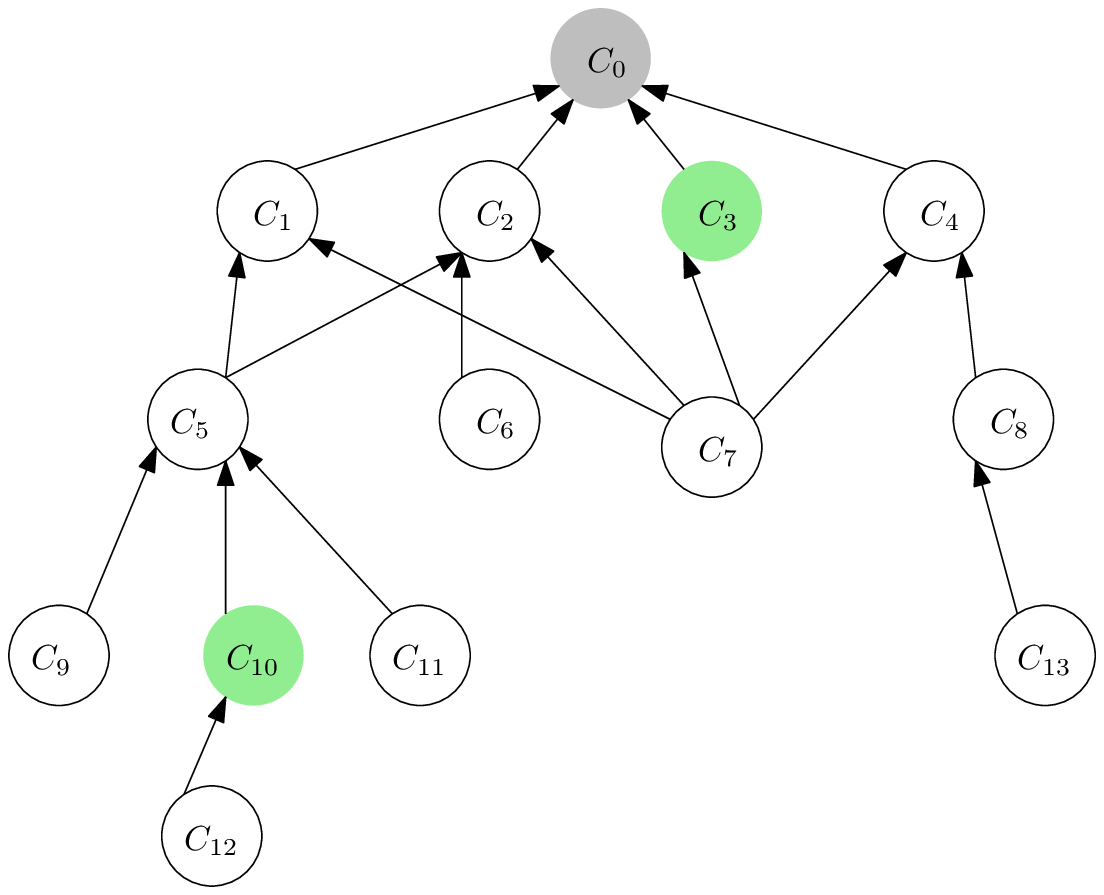}
 \caption{An example ontology of hyponymy-hypernymy relationship.}
 \end{center}
\end{figure}
\noindent to the $Fig.~2$, there are two concepts $c_3$ and $c_{10}$ which are at different depths. $c_3$ at depth 1 and $c_{10}$ at depth 3. 
Both has single leaf node. Now leaf node of $c_3$ has six numbers of subsumers and leaf node of $c_{10}$ also has six numbers 
of subsumers then $c_3$ and $c_{10}$ will have same IC. But it should not happen as both the concepts are at different depth and as we know 
that the more deep we proceed in any ontology the more greater IC the node possess\cite{52}. Along with this, topological structure of subsumers of both the
leaves are also different.\\ 
\begin{figure}[!htb]
 \begin{center}
\includegraphics[scale=.55]{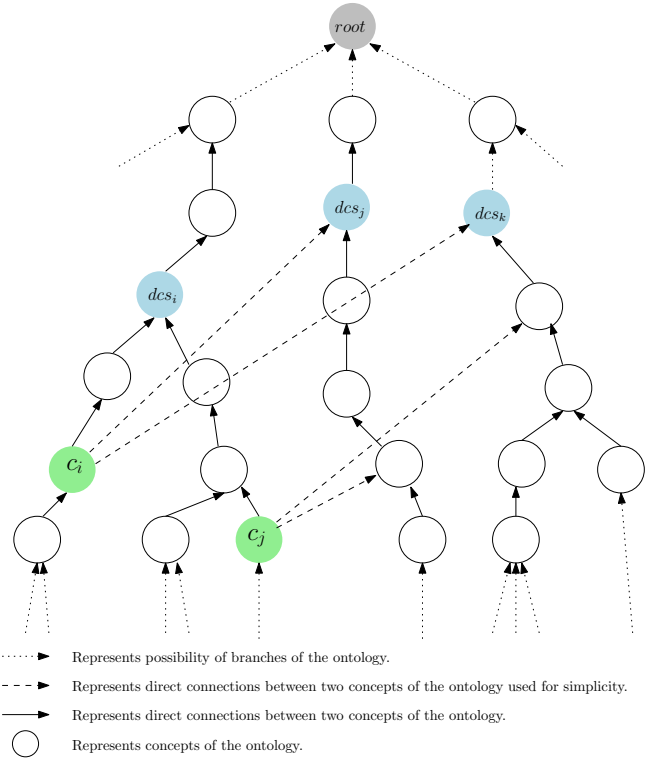}
 \caption{A partial view of an example ontology with hyponymy-hypernymy relationship and members of set $DCS(c_i, c_j)$.}
 \end{center}
\end{figure}
\indent It is quite evident that without an effective semantic similarity calculator, we can not determine semantic similarity between two concepts more accurately
based on semantics. 
We find some new aspects which are missing in existing competing methods to incorporate into a semantic similarity calculation model. 
$DCS(c_i, c_j)$ is one of them. $DCS(c_i, c_j)$ is defined in definition 8 and Algorithm 1 shows the procedure to find $DCS(c_i, c_j)$. 

To overcome the above mentioned issues we need to design a novel intrinsic IC calculation model along with a novel semantic similarity calculation model
which can generate better \emph{correlation coefficient} value when applied to benchmark data. That is why, we propose a new intrinsic IC calculation model
based on several structural aspects of the ontology. These aspects are discussed in section 5. Apart from IC calculation technique, we also propose a novel semantic similarity calculator in this paper.

\section{System model and Metric used} 
The problem can be efficiently modeled by the notion of \emph{discrete mathematics}\cite{51}. An Ontology $O$ is
a connected graph G(V, E) where vertex set $V$ represents set of concepts, i.e. $ V= \{c_0, c_1, c_2, \ldots, c_n\}$, and 
edge set $E$ represents relation R between concepts.
So, $R\subseteq V \times V$ where $R\in {Hyponym}\vee{Hypernym}$. Along with this, R is transitive and neither reflexive nor
symmetric in nature. IC of a concept is denoted as $IC(c)$ $\mid$ $c$ is a concept. $IC(c)\in[0,1]$.
IC(root) in any topology is zero. Maximum value of IC could be 1 and the concept having IC value 1 must be a leaf node. $Sim(c_i, c_j)$ denotes the semantic similarity calculator for concepts $c_i$ and $c_j$. $Sim(c_i,c_j)\in[0,1]$.
\\
\indent We use correlation coefficient \cite{20} as metric to check performance of our proposed information theoretic framework. For finding 
correlation coefficient we need two sets of data, $X~and~Y$. When two sets of data are strongly linked together then we say they are highly 
correlated. correlation coefficient  value ranges between -1 to +1. The formula for calculating correlation is as follows:
\begin{equation}
\label{eqn:correlation}
 r_{xy}=\frac{\Sigma_{i=1}^{n}(x_{i}-\bar{x})(y_i-\bar{y})}{\sqrt{\Sigma_{i=1}^{n}(x_{i}-\bar{x})^2\Sigma_{i=1}^{n}(y_i-\bar{y})^2}}
\end{equation}
where, $X=(x_1, x_2, \ldots, x_n)$ and $Y=(y_1, y_2, \ldots, y_n)$.  $(x_{i}-\bar{x})$ is the difference between each term of set $X$ and mean of set $X$ and 
$(y_i-\bar{y})$ is the difference between each term of set $Y$ and mean of set $Y$. 
We have compared this correlation value with correlation values of several
existing semantic similarity calculation model to measure performance of our IC calculation models along with our semantic similarity model.
In the current work, we round up all the correlation coefficient values up to two decimal places.

\section{Proposed Solution}
In this section we present our model for calculating IC intrinsically and model for finding semantic similarity based on calculated IC. Before going 
to discuss our solution we present some definition which we consider in our framework throughout. Following this we describe our proposed solution.
\subsection{Definitions}
\newtheorem*{Definition}{Definition 1}
\begin{Definition}
$Hyponyms(c)=\{ a \mid a \in V \wedge \forall$ $a$, $a$ $\preceq$ $c$ $\}$, i.e. 
all the concepts which are subsumed by the concept $c$.
\end{Definition}
\newtheorem*{Definition2}{Definition 2}
\begin{Definition2}
$ Instance~hyponyms(c)=$ It represents specific(usually real world) instance of node $c$. e.g. instance hyponym of mountain is Evererst.
\end{Definition2}
\newtheorem*{Definition3}{Definition 3}
\begin{Definition3}
$ deep(c)=$ It represents minimum distance of $c$ from $root$ node of the ontology.
\end{Definition3}
\newtheorem*{Definition4}{Definition 4}
\begin{Definition4}
 $Subsumers(c)=\{a \in V,~c \in V \mid c \preceq a\}\cup\{c\}$, $c \preceq $ $a$ means that $c$ is a hierarchical specialization of $a$.
where, $V$ is the set of concepts in the ontology.
\end{Definition4}
\newtheorem*{Definition5}{Definition 5}
\begin{Definition5}
$ Leaves(c)=$ $\{l \in V,~c \in V \mid l\in hyponyms(c)$ $\Lambda$ $l$ is a leaf$\}$,
where, $V$ be the set of concepts of the ontology, $l$ is a leaf iff $Hyponyms(l) = \phi$ $\Lambda$ $Instance~hyponyms(l) = \phi$.
\end{Definition5}
\newtheorem*{Definition6}{Definition 6}
\begin{Definition6}
$nmih(c) =$ number of subsumers which are directly connected to the concept  $c$ by a single link in the ontology.
\end{Definition6}
\newtheorem*{Definition7}{Definition 7}
\begin{Definition7}
$Multiple~Inheritance=$ If a concept $c$ has more than one direct subsumers, then we can say the concept $c$ has multiple inheritance.
\end{Definition7}
\newtheorem*{Definition8}{Definition 8}
\begin{Definition8}
$ DCS(c_i, c_j)=$ It stands for Disjoint Common Subsumers. It represents set of those nodes which subsume both $c_i$ and $c_j$ but none of them are related to each other by
Hyponym-Hypernym relation. Members of the set $DCS(c_i, c_j)$ are denoted as $dcs$. \\
\indent Pictorially $dcs$ is shown in $Fig.~3$. According to $Fig.~3$ there are three $dcs$ present for concepts $c_i$ and $c_j$. Those are coloured as light blue. 
Procedure for calculating $dcs$ is formally written in Algorithm 1. First, consider four empty sets. Those are, $DCS(c_i,c_j)$, $DCS_{suspect}(c_i,c_j)$,
$S_i$, $S_j$. Secondly, store all the subsumers of concept $c_i$ and $c_j$ in $S_i$ and $S_j$ respectively. Find the intersection of these two sets $S_i$ and $S_j$.
Store them in $DCS_{suspect}(c_i,c_j)$. Perform a sorting operation on the set $DCS_{suspect}(c_i,c_j)$ in descending order based on depth of each element of that set.
Pick the largest element as first $dcs$ from the set $DCS_{suspect}(c_i,c_j)$ and assign to $DCS(c_i,c_j)$. Discard the element from $DCS_{suspect}(c_i,c_j)$.
Pick the next largest element $x$ from $DCS_{suspect}(c_i,c_j)$ and check whether any of the element of $DCS(c_i,c_j)$ is hyponym of $x$. If none of the elements 
of $DCS(c_i,c_j)$ are hyponym of $x$ then add this $x$ into set $DCS(c_i,c_j)$ and discard it from $DCS_{suspect}(c_i,c_j)$. 
Otherwise discard $x$ from $DCS_{suspect}(c_i,c_j)$ only, and 
do not add to $DCS(c_i,c_j)$. Do it until all the elements of $DCS_{suspect}(c_i,c_j)$ are checked.
\end{Definition8}

\begin{algorithm}[H]
\scriptsize
\caption{Calculating $DCS(c_i,c_j)$}
\algrenewcommand{\alglinenumber}[1]{\bfseries Step {#1}.}
\renewcommand{\Statex}{\item[\hphantom{\bfseries Step \arabic{ALG@line}.}]}
\begin{algorithmic}[1]
\State Start
\State Initialize set \textit{$DCS(c_i,c_j)$ $\leftarrow$ $\phi$}, set \textit{$DCS_{suspect}(c_i,c_j)$ $\leftarrow$ $\phi$}, 
set \textit{$S_i$ $\leftarrow$ $\phi$}, set \textit{$S_j$ $\leftarrow$ $\phi$}.
\State For concept $c_i$ add all Subsumers($c_i$) to set $S_i$  and for concept $c_j$ add all Subsumers($c_j$) to set $S_j$.
\State set $DCS_{suspect}(c_i,c_j)$ $\leftarrow$ $S_i$ $\cap$ $S_j$.
\State Sort elements of set $DCS_{suspect}(c_i,c_j)$ in descending order based on depth of each elements.
\State Assign the largest element of the set $DCS_{suspect}(c_i,c_j)$ as first $dcs$ to  $DCS(c_i,c_j)$.
\State set $DCS_{suspect}(c_i,c_j)$ $\leftarrow$ $DCS_{suspect}(c_i,c_j)$ - $DCS(c_i,c_j)$.
\State for each element $x$ in $DCS_{suspect}(c_i,c_j)$
\Statex ~~~if any member of $DCS(c_i,c_j)$ is in the set $Hyponym(x)$ 
\Statex	~~~~~~discard that node $x$ from $DCS_{suspect}(c_i,c_j)$
\Statex	~~~else add $x$ to $DCS(c_i,c_j)$ and repeat Step 7.
\State End
\end{algorithmic}
\label{algo5}
\end{algorithm}

\subsection{Our approach}
\subsubsection{Our intrinsic IC calculation model:}
Designing an IC calculation model to quantify information for each and every concepts of the ontology in an intrinsic way is really a hard challenge.
In this section we present a new intrinsic model to compute IC of a concept from WordNet. It is quite obvious that the more
we consider structural aspects of the ontology, the more are the chances of getting accurate results in finding IC. Depth of any concept is one of 
the important factors in calculating IC. That is why leaves at different depth should not have same IC value.
Again two nodes having same hyponym structures and staying in the same depth but having different number of subsumers should not have same IC value.
Also nodes having different depth 
but same number of subsumers should not have same IC. So, another key factor is number of multiple inheritance the concepts have.
According  to our observation, 
IC of any concept is inversely proportional to the number of multiple inheritance. Apart from this, number of subsumers of any concept is directly 
proportional with its IC value. Number of hyponyms are also important factor as it decides generality of the concept and it is inversely proportional
with the IC. Based on this notion and considering several other topological factors, we propose a new intrinsic IC calculation model in this paper as follows:
\begin{align}
\label{eqn:Ouric}
IC_{our}(c)=\frac{\log(deep(c)+1)}{\log(deep_{max}+1)} \nonumber \\  \times (1-\log(\frac{\frac{leaves(c)\times(nmih(c))}{leaves_{max}}}{subsumer(c)}+1)) \\ 
\times (1- \frac{\log((\Sigma_{a\in hypo(c)} \frac{1}{deep(a)}) + 1)}{\log(node_{max})})\nonumber
\end{align}
where, $deep(c)$ is depth of concept $c$, $deep_{max}$ is maximum depth of the Ontology (in our case WordNet version 3.0), $leaves(c)$ is number of leaves of 
concept $c$, $nmih(c)$ is number of multiple inheritances of concept $c$, $leaves_{max}$ is maximum number of leaves of the ontology, $subsumer(c)$ 
is number of subsumers of concept $c$, $hypo(c)$ is number of hyponyms of concept $c$, $node_{max}$ is maximum number of nodes in the ontology.
In our experimental setup when ever we need to use $hypo(c)$, we include all the members of Instance~hyponyms(c) in the set of $hypo(c)$.

\subsubsection{Our Semantic Similarity calculation model based on IC:}
After getting the IC of concepts, it is the responsibility of a semantic similarity calculator to measure semantic similarity accurately.
So, we also concentrate on desining a novel semantic similarity calculator and propose a new model which can act based on our IC model as well
as others'. 
In designing our own model we consider pure information theoretic perspective and some structural aspects of the underlying ontology. 
We introduce Disjoint Common Subsumer ($DCS$). We give weightage to this $DCS$ structural aspect of ontology and also consider some ratio factors for formulating our model as follows:
\begin{equation}
\label{eqn:Oursim}
 Sim_{our}(c_i, c_j)=\frac{\sum_{r=1}^{m} \{\frac{IC(dcs_r)}{IC(c_i) +1}+\frac{IC(dcs_r)}{IC(c_j)+1}\}} {m}
\end{equation}
where, $m$ denotes size of the set $DCS(c_i, c_j)$ for concepts $c_i$ and $c_j$. Every $dcs$ have some vital contribution for deciding 
semantic similarity between two concepts. Each member of $DCS(c_i, c_j)$ actually holds the semantic similarity factor of any two concepts $c_i$ and $c_j$. 
It gives a new semantic-dimension to the concepts under consideration and enhance the scope of each concept to 
be semantically similar to each other. We consider this dimension also.
Again, the more distant, concepts $c_i$ and $c_j$ are from $dcs_r$, the more dissimilar the two concepts would be. Because, a 
concept gets more concreteness while it goes deeper from its root in the ontology. So, when the concept goes deeper from its $dcs$, it gets more IC value. 
Concreteness also gets higher of that concept. The ratio factor $1/\{IC(c_j)+1\}$ gets smaller due to increasing IC value and it also becomes a factor for 
reducing the overall semantic similarity between concepts. So, we consider 
$dcs$ with individual concept's IC ratio model and propose \emph{equation \ref{eqn:Oursim}}.

\indent It is important to note that in WordNet, several words have multiple senses(synsets), i.e. $W=\{s_1, s_2, s_3, \ldots, s_n\} $. 
These words are called polysemic words.
In such cases we have computed semantic similarity based on the following formula:
\begin{equation}
\label{eqn:polysemic}
 Sim(W_x, W_y)=\max(sim(s_{xi}, s_{yj}))
\end{equation}
where, $s_{xi}$, $s_{yj}$ are senses (i.e. concepts or synsets in the WordNet ontology) of polysemic words $W_x$ and $W_y$ respectively.
We use a linear transformation of Jiang and Conrath distance formula (\emph{eq. \ref{eqn:Jiangsim}}) to a semantic similarity function 
which has been used by Seco et al.:
\begin{equation}
 Sim(c_i, c_j)= 1-(\frac{IC(c_i) + IC(c_j) - 2 \times Sim_{res}(c_i, c_j)}{2})
\end{equation}

\section{Experiments}
In this section, we describe first the experimental setup in subsection $6.1$. We describe results gained by our experiment along with discussions
regarding our results in subsection $6.2$.
\begin{table}[H]
  \caption{Shows competing methods used in our paper.}
  \label{T1competingmethods}
 \begin{center}
 \fontsize{10}{0} \scalebox{1.1}
{
 \begin{tabular}{p{4cm} p{2.5cm} } 
 \hline
  Approach & Description \\
 \hline\hline
 Resnik\cite{1} & IC(corpora-based)\\
 Lin\cite{3} & IC(corpora-based)\\
 Jiang and Conrath\cite{2} & IC(corpora-based)\\
 Seco et al.\cite{5} & IC (intrinsic)\\
 Zhou et al.\cite{7} & IC (intrinsic)\\
 S\'{a}nchez et al.2011\cite{6} & IC (intrinsic)\\
 S\'{a}nchez et al.2012\cite{4} & IC (intrinsic)\\
 Montserrat Batet et al.\cite{47}  & IC (intrinsic)\\
 Meng et al.\cite{8}  & IC (intrinsic)\\
 Qingbo et al.\cite{50}  & IC (intrinsic)\\
 Pirr\'{o} et al.\cite{48} & IC (intrinsic)\\
 Rada et al.\cite{10} & Edge-counting\\
 Wu and Palmer\cite{11} & Edge-counting\\
 Leacock and Chodorow\cite{12} & Edge-counting\\
 Li et al.\cite{13} & Edge-counting\\
 Rodriguez and Egenhofe\cite{14} & Feature-based\\
 Tversky\cite{15} & Feature-based\\
 Petrakis et al.\cite{16} & Feature-based\\
 Aida Valls et al.\cite{46} & Feature-based\\
 Bollegala et al.\cite{17} & Distributional\\
 Chen et al.\cite{8} & Distributional\\
 Sahami and Heilman\cite{19} & Distributional\\
 Gledsone et al.\cite{45} & Distributional  \\
 Danushka Bollegala et al.\cite{44} & WebSnippest and Page-count based\\
 
      \hline
\end{tabular}
}
\end{center}
\end{table}
\subsection{Experiment Setup}
Our background ontology is WordNet version 3.0. It has more than 1,00,000 English concepts and are organized in a very meaningful
way according to human cognition. There are four different topologies under the whole WordNet version 3.0.
Those are noun, verb, adjective and adverb. In our experiment, we consider only the noun. There are many pollysemic words under WordNet. Each word is compossed of several synsets.
Each synset represents a distinct sense. Every synset has a set of synonyms. Synsets are interlinked
by means of conceptual-semantic and lexical relations.
For data set we have used Miller and Charles'\cite{9} benchmark data and Wordism similarity
goldstandard\cite{4} benchmark data set. Miller and Charles have assessed 30 noun pairs by 38 students based on a scale of 0 (semantically unrelated) 
to 4 (highly synonymous). Wordism similarity goldstandard is a part of the well-known WordSim353 \cite{53} test collection. It consists of a set of 203
word-pairs. But as we consider only noun-pairs for our measure, only 201 word-pairs from that wordism data set are taken under consideration. Remaining two
word-pairs are not nouns.
These benchmark data is considered as de facto data due to extensive use in many related works to asses performance of their models.
We use NLTK (Natural Language Toolkit) \cite{39} which is a leading platform to access WordNet version 3.0. 
We use python as implementation language. NLTK provides easy-to-use interfaces to cover 50 corpora and lexical resources such as WordNet, 
along with a suite of text processing libraries for classification, parsing, tokenization, semantic reasoning and so on.
For fair comparison of our proposed information theoretic semantic similarity framework with state of the art methods we have implemented all the intrinsic IC based competing methods of
\emph{TABLE \ref{T1competingmethods}} in our experimental setup.
In order to evaluate our IC model we use three different classical semantic similarity models along with Pirr\'{o} et al. mentioned in section 2 and our
proposed semantic similarity model. 
Proposed semantic similarity calculator has also been evaluated using several IC models including our IC model.

\subsection{Results and Evaluation}
In this subsection, we show all the experimental results and provide a thorough comparison of our proposed semantic similarity framework
with the existing competing methods. 
Individual semantic similarity scores, obtained by our IC calculation approach when applied to 
Resnik, Lin, J\&C and Pirr\'{o} semantic similarity models for each 30 noun pairs of M\&C benchmark data, is shown in \emph{TABLE \ref{T2similarityscore30data}}.
Similarity scores obtained by our proposed semantic similarity measure based on our proposed IC calculation model and IC model proposed by Meng et al. 
for M\&C benchmark data set, has also been showcased in \emph{TABLE \ref{T2similarityscore30data}}. Similarity scores, obtained by our IC calculation approach when applied to 
Resnik, Lin, J\&C and Pirr\'{o} semantic similarity models for Wordism similarity goldstandard data set, is shown in 
\emph{TABLE 1} to \emph{TABLE 4} in Appendix A.
Similarity scores obtained by our proposed similarity measures with our proposed IC model and IC model proposed by Meng et al.
for Wordism similarity goldstandard data set, has also been showcased in \emph{TABLE 1} to \emph{TABLE 4} in Appendix A.

\begin{table*}[!htb]
 \caption{Correlations and semantic similarity scores obtained by our intrinsic IC calculation approach (\emph{eq.\ref{eqn:Ouric}}) when applied to Resnik's,
 Lin's, Jiang and Conarth's, Pirro's and Our's semantic similarity measures for each noun pairs of the Miller and Charles' benchmark data-set of 30 noun pairs.}
\label{T2similarityscore30data}
  \begin{center}
  \fontsize{10}{0} \scalebox{1}{
 \begin{tabular}{||p{3.5cm} p{1.5cm} p{1.5cm} p{2cm} p{1.5cm} p{1.5cm} p{1.5cm}||} 
 \hline
 Noun pairs &Resnik(IC computed as eq.\eqref{eqn:Ouric}) & Lin(IC computed as eq.\eqref{eqn:Ouric}) & Jiang\&Conarth (IC computed as eq.\eqref{eqn:Ouric}) & Pirr\'{o}(IC computed as eq.\eqref{eqn:Ouric})& Our Sim(IC computed as eq.\eqref{eqn:Ouric})& Our Sim(IC computed as Meng)\\ 
 
 \hline\hline
 1.Car-automobile &  0.704
 & 1.000 & 1.000 & 1.000 & 0.826 & 0.815\\ 
 \hline
 2.Gem-jewel & 0.701
 & 1.000 & 1.000 & 1.000 & 0.824 & 0.806\\
 \hline
 3.Journey-voyage & 0.663
 & 0.905 & 0.930 & 0.826 & 0.766 & 0.753
\\
 \hline
 4.Boy-lad & 0.537
 & 0.897 & 0.938 & 0.813  & 0.673 & 0.633
\\
 \hline
 5.Coast- shore & 0.491
&  0.935 & 0.966 & 0.879  & 0.645 & 0.587
\\ 
 \hline
 6.Asylum-madhouse & 0.775
 &  0.975 & 0.980 & 0.952 & 0.864 & 0.849
\\
 \hline
  7.Magician-wizard & 0.564
 &1.000  & 1.000 & 1.000 &0.721 & 0.680\\
\hline
 8.Midday-noon & 0.782
 & 1.000 & 1.000 & 1.000 &0.877 & 0.863\\
\hline
 9.Furnace-stove & 0.199
 & 0.274 & 0.473 & 0.159 &0.231 & 0.207
\\
\hline
10.Food-fruit & 0.128
 &  0.190 &0.649 & 0.105 &0.153 & 0.103
 \\
\hline
11.Bird-cock & 0.488
 & 0.753  & 0.840 & 0.604 &0.598 & 0.587
\\
\hline
12.Bird-crane & 0.488
 & 0.721  & 0.811 & 0.564 & 0.589 & 0.578	
\\
\hline
13.Tool-implement & 0.405
& 0.913  &0.961 & 0.840 & 0.561 & 0.535
 \\
\hline
14.Brother-monk & 0.618
 & 0.933  & 0.956 & 0.876 & 0.744 & 0.714
\\
\hline
15.Crane-implement& 0.262
 & 0.463  & 0.696 & 0.302 &0.339 & 0.316
\\
\hline
16.Lad-brother &0.161
  & 0.268  & 0.561 & 0.155 &0.201 & 0.170
\\
\hline
17.Journey-car & 0.000
 &  0.000 &0.337 & 0.000 &0.000 & 0.000
 \\
\hline
18.Monk-oracle & 0.161
 & 0.248  &0.511 & 0.141 &0.195 & 0.163
 \\
\hline
19.Cemetery-woodland & 0.090
 & 0.145  &0.469 & 0.078 &0.111 & 0.078
 \\
\hline
20.Food-rooster & 0.048
 & 0.082  & 0.457 & 0.042  &0.063 & 0.000
\\
\hline
21.Coast-hill & 0.301
 & 0.553  &0.756 &  0.382 &0.390 & 0.325
 \\
\hline
22.Forest-graveyard & 0.090
 & 0.145  & 0.469 & 0.078 &0.111 & 0.078
\\
\hline
23.Shore-woodland &  0.090
& 0.181  &0.593 & 0.099 &0.120 & 0.085
 \\
\hline
24.Monk-slave & 0.161
 & 0.291  & 0.607 &  0.170 &0.208 & 0.176
\\
\hline
25.Coast-forest & 0.090
 &0.169 & 0.559 & 0.092 &0.117 & 0.083
\\
\hline
26.Lad-wizard &  0.161
& 0.277  & 0.579 & 0.160  &0.203 & 0.172
\\
\hline
27.Chord-smile & 0.155
 &0.227  &0.471 & 0.128 & 0.184 & 0.122
 \\
\hline
28.Glass-magician &  0.123
& 0.202 &0.530 & 0.112 &0.153 & 0.131
 \\
\hline
29.Noon-string & 0.050
 & 0.077 & 0.396 & 0.040 &0.061 & 0.000
\\
\hline
30.Rooster-voyage & 0.000
 & 0.000 &0.153 & 0.000  &0.000 & 0.000
 \\
\hline
\rowcolor{lightgray}
CORRELATION: & 0.86
 & 0.86  &0.84 &  0.86 &0.86 & 0.87
 \\
 \hline
\end{tabular}
}
\end{center}
\end{table*} 

\begin{figure}[H]
 \begin{center}
\includegraphics[scale=.22]{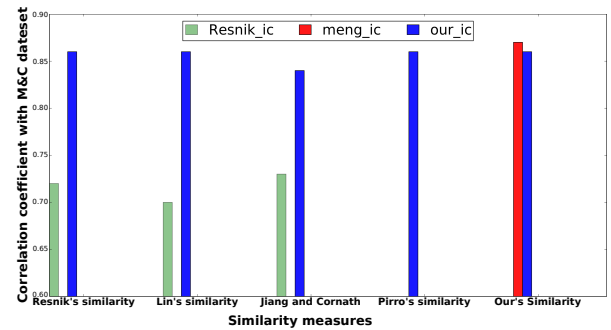}
 \caption{Correlation coefficient values obtained by our proposed framework vs. state of the art semantic similarity measures based on corpora based IC calculation technique.}
 \label{plot3corporavsintrinsic}
 \end{center}
\end{figure}

\begin{figure}[htbp]
 \begin{center}
\includegraphics[scale=.22]{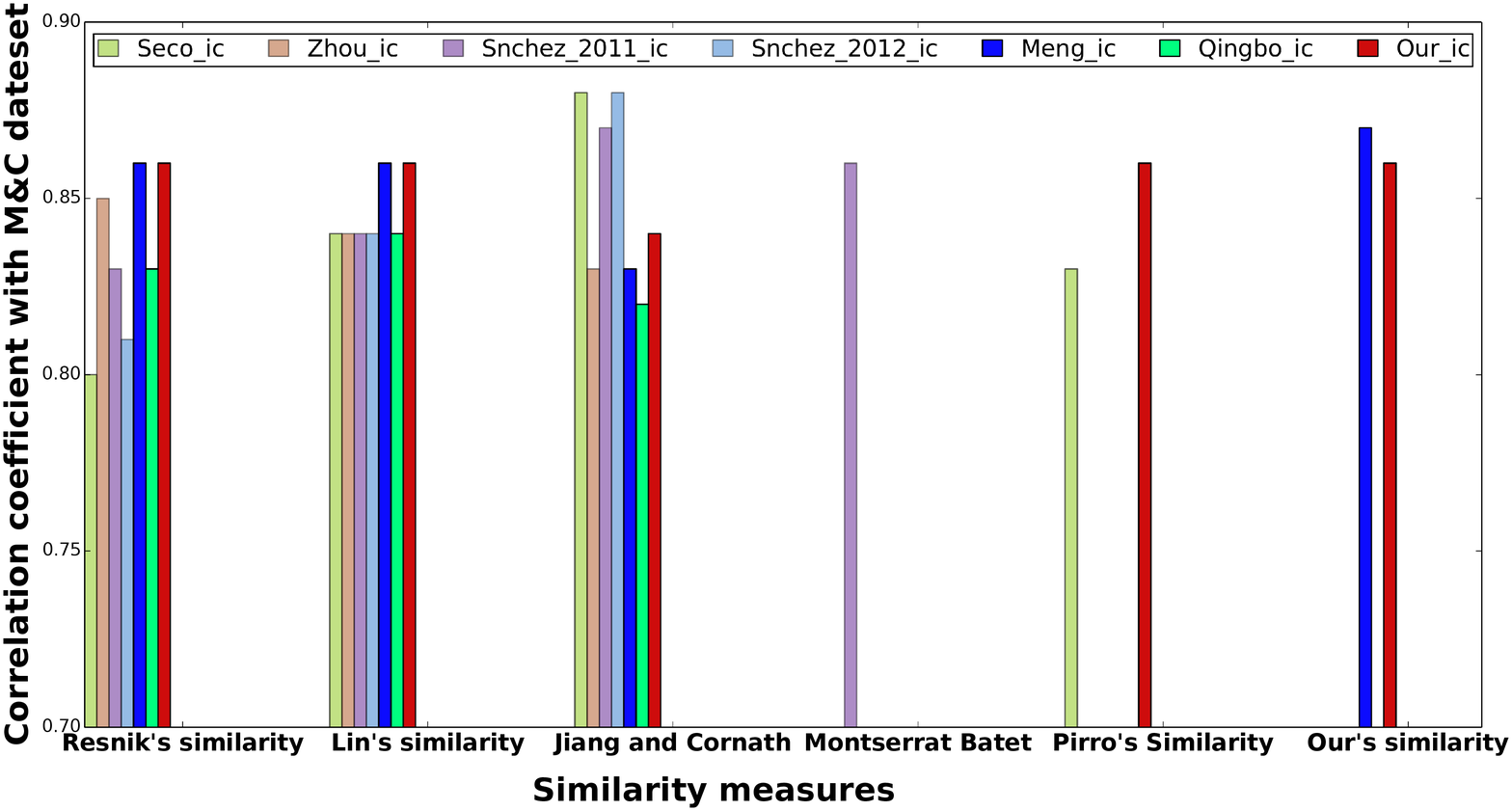}
 \caption{ Correlation coefficient values of our proposed framework  vs. state of the art 
 intrinsic IC based semantic similarity measures based on M\&C dada set.}
 \label{plot4intrinsicvsintrinsic30data}
 \end{center}
\end{figure}

We first compare our intrinsic IC model with some corpora based information theoretic approach. For this we select Resnik, Lin, and J\&C semantic similarity measures
with corpora based IC calculation techniques. \emph{TABLE \ref{3corporabasedapproach}} shows correlation coefficient values obtained by 
these similarity calculation techniques and proposed intrinsic IC-model based semantic similarity measure. 
In \emph{TABLE \ref{3corporabasedapproach}} the last column shows correlation values with respect to Miller and Charles' benchmark data.
We show a bar-chart for this particular experiment depicted in $Fig.~\ref{plot3corporavsintrinsic}$. $X$ axis shows different semantic similarity measures. $Y$ axis shows 
correlation coefficient value with M\&C data set. This bar-chart compares correlation coefficient values of our proposed intrinsic IC
based semantic similarity measure versus state of the art semantic
similarity measures based on corpora based IC calculation
technique.
\emph{TABLE \ref{3corporabasedapproach}} and $Fig.~\ref{plot3corporavsintrinsic}$ clearly show that 
our proposed intrinsic IC model with our proposed semantic similarity 
model and with the Pirro's semantic similarity model give far better results than corpora based IC calculation techniques.  
Our semantic similarity model with Meng's intrinsic IC model also gives better result than any of the corpora based IC calculation techniques.
\\
\indent We evaluate our proposed IC model and proposed semantic similarity model with existing intrinsic IC based semantic similarity models. 
For this evaluation we select all of the competing intrinsic IC based semantic similarity methods enlisted in
\emph{TABLE \ref{T1competingmethods}}.
Based on our IC model and 30 noun pairs, we get correlation coefficient
value of 0.86 using Resnik semantic similarity measure. Using our IC with Lin semantic similarity model, we get 0.86 correlation coefficient 
value and using Jiang and Conrath semantic similarity measure we get 0.84 correlation coefficient value. 
Our IC with Pirr\'{o} et al. and our IC with our semantic similarity model also gives 0.86 correlation value. All the correlation coefficient
values found by our IC model are within the upper limit for 30 noun pairs which is 0.88\cite{1} and the upper bound for
28 noun pairs is 0.90 \cite{1}. Correlation values by Resnik, Lin, Pirr\'{o} and our semantic similarity measures based on our 
intrinsic IC calculation model are very close to the upper bound. Results obtained by Resnik, Lin, Pirr\'{o} and our semantic similarity model
based on our IC model are similar. On the contrary, in other's IC models there are significant gap exists between these measures. 
Also Resnik shows very poor results based on other IC models. Whereas based on our IC model with 30 noun pairs and 201 noun pairs, 
we get best correlation coefficient found ever by Resnik means.
Our semantic similarity model with Meng et al. IC calculation model 
shows more significant results than existing IC based measures. It gives 0.87 correlation with $M\&C$ benchmark data.
$Fig.~\ref{plot4intrinsicvsintrinsic30data}$ shows the comparison of
correlation coefficient values of our proposed semantic
similarity measure embedded with Meng et al. IC model and
our's versus state of the art intrinsic IC based semantic similarity
measures. It also shows correlation coefficient values of some of the existing similarity measures embedded 
with our intrinsic IC model versus state of the art intrinsic IC
based semantic similarity measures. 
$X$ axis shows different semantic similarity measures. $Y$ axis shows 
correlation coefficient value with M\&C data set. \emph{TABLE \ref{4intrinsicapproach30data}} shows detailed correlation coefficient values 
of different semantic similarity calculation techniques based on different intrinsic IC models
\begin{table}[H]
  \caption{Shows correlation coefficient values of three classical information theoretic semantic similarity measures based on corpora based IC calculation
  technique and correlation values obtained by our proposed framework.}
  \label{3corporabasedapproach}
 \begin{center}
\fontsize{10}{0}  \scalebox{1}{
 \begin{tabular}{p{3cm} p{1.7cm} p{.9cm} p{1.1cm}} 
 \hline
  Similarity measures&Type & No. of&Correlation\\ [0.5ex] 
  (with IC model)&   &noun pairs&with M\&C\\
 \hline\hline
 Resnik (IC ~\\computed as Resnik\cite{1}) & IC(corpora-based)  & 28  & 0.72  \\ 
   Lin  (IC ~\\ computed as Resnik\cite{1}) & IC(corpora-based) & 28  & 0.70   \\
    Jiang and Conrath ~\\ (IC computed as Resnik\cite{1}) & IC(corpora-based) & 28  & 0.73   \\
   \hline 
     \rowcolor{lightgray}
     Resnik (IC computed & & &\\ 
     \rowcolor{lightgray} by our model) & IC (intrinsic)  & 30  & 0.86    \\
     \rowcolor{lightgray}
     Lin (IC computed & & &\\ 
     \rowcolor{lightgray} by our model) &  IC (intrinsic)& 30 & 0.86\\
    \rowcolor{lightgray}
    Jiang and Conarth & & &\\ 
    \rowcolor{lightgray}(IC computed by our model) &  IC (intrinsic)& 30 & 0.84 \\
     \rowcolor{lightgray}
    Pirr\'{o} & & &\\ 
    \rowcolor{lightgray}(IC computed by our model) &  IC (intrinsic)& 30 & 0.86 \\
     \rowcolor{lightgray}
     Our\_Sim (IC computed & & &\\ 
     \rowcolor{lightgray} by our model) & IC (intrinsic)  & 30  &    0.86  \\
      \rowcolor{lightgray}
    Our\_Sim & & &\\ 
    \rowcolor{lightgray}(IC computed by Meng et al. model) &  IC (intrinsic)& 30 & 0.87 \\
     [1ex] 
      \hline
\end{tabular}
}
\end{center}
\end{table}

\begin{figure}[htbp]
 \begin{center}
\includegraphics[scale=.22]{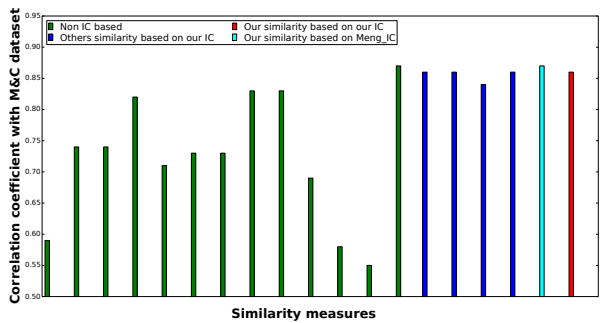}
 \caption{Correlation coefficient values of our proposed framework vs. non-IC based state of the art semantic similarity measures.}
 \label{plot5nonicvsintrinsic}
 \end{center}
\end{figure}

\noindent. \emph{TABLE \ref{4intrinsicapproach30data}} also shows results obtained by our proposed framework.
In this \emph{TABLE \ref{4intrinsicapproach30data}} the last columns shows correlation values with respect to Miller and Charles'
benchmark data.\\
\indent There are several non-IC based methods available for measuring semantic similarity between concepts. So, we compare proposed IC model and semantic similarity 
model with these non-IC based models. \emph{TABLE \ref{5nonicbasedmethods}} shows correlation coefficient values of different non-IC based semantic similarity
calculation techniques and proposed IC model along with proposed semantic similarity model. In this \emph{TABLE \ref{5nonicbasedmethods}} the last column shows correlation values with 
respect to Miller and Charles' benchmark data. It is quite evident from the \emph{TABLE \ref{5nonicbasedmethods}} that, none of the non-IC based 
semantic similarity measures out perform our IC model along with our semantic similarity model.
$Fig.~\ref{plot5nonicvsintrinsic}$ compares correlation coefficient values of our proposed semantic
similarity measure embedded with IC model of Meng and our's
versus non-IC based state of the art semantic similarity measures. It 
also shows correlation coefficient values of some of the existing similarity measures embedded with
our intrinsic IC model versus non-IC based state of the art semantic
similarity measures. $X$ axis shows different semantic similarity measures. $Y$ axis shows 
correlation coefficient value with M\&C data set.
\indent \\
\indent To evaluate our proposed IC calculator and similarity model more precisely, we use an additional recent benchmark 
data set, named Wordism similarity goldstandard.

\begin{figure}[htbp]
 \begin{center}
\includegraphics[scale=.22]{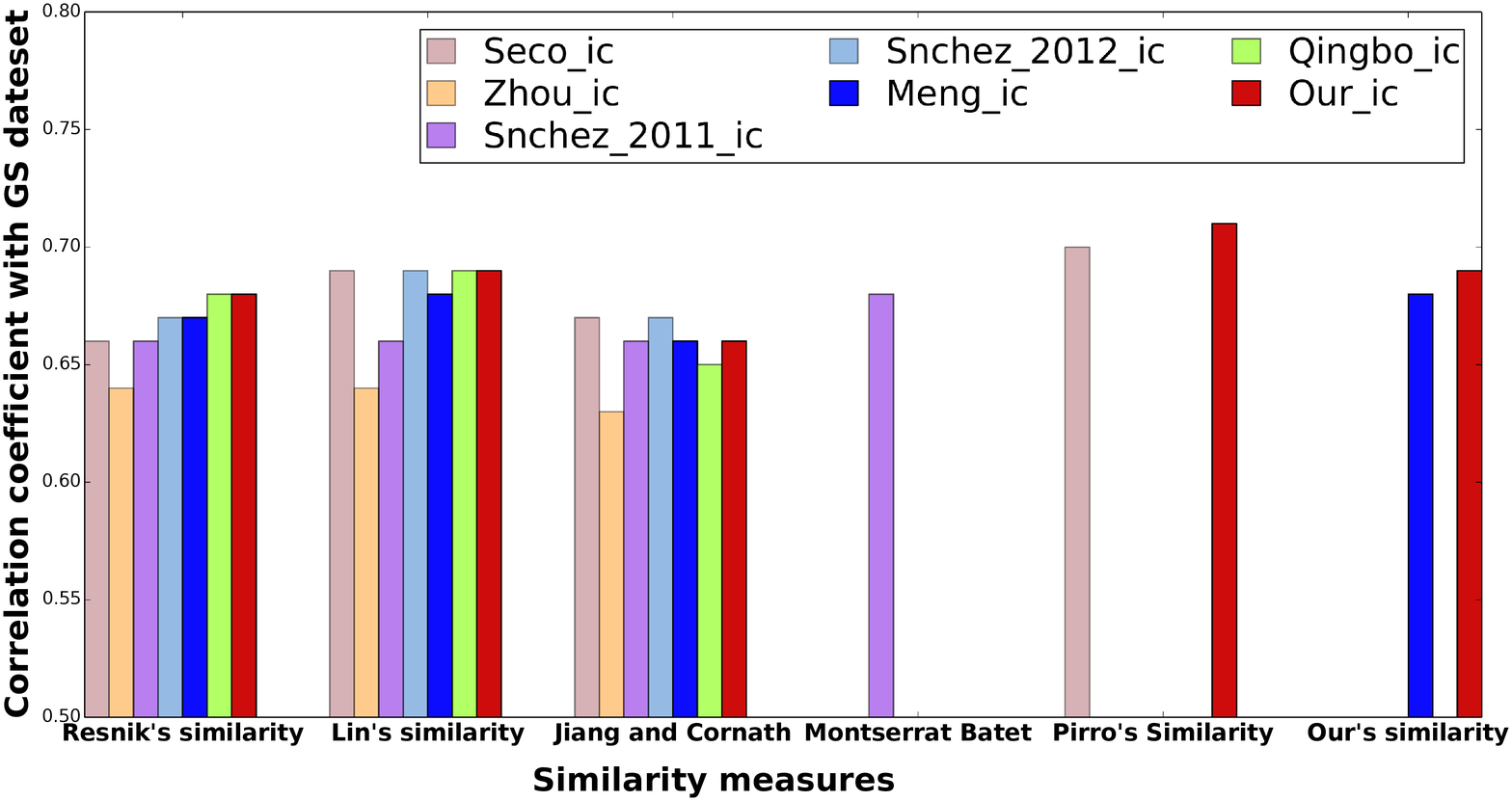}
 \caption{Correlation coefficient values of our proposed framework vs. state of the art 
 intrinsic IC based semantic similarity measures
 based on goldstandard data set.}
 \label{plot6intrinsicvsintrinsic201data}
 \end{center}
\end{figure}

\begin{table}[htbp]
  \caption{Shows correlation coefficient values for different semantic similarity measures based on different intrinsic IC model
  and correlation values obtained by our proposed framework.}
    \label{4intrinsicapproach30data}
 \begin{center}
 \fontsize{10}{0} \scalebox{1}{
 \begin{tabular}{p{3.2cm} p{1.7cm} p{.9cm} p{1.1cm}} 
 \hline
  Similarity measures&Type & No. of&Correlation\\ [0.5ex] 
  (with IC model)&   &noun pairs&with M\&C\\
 \hline\hline
     Resnik (IC ~\\computed as Seco et al.\cite{5}) & IC (intrinsic) & 30 & 0.80  \\
     Lin (IC ~\\ computed as Seco et al.\cite{5}) &  IC (intrinsic)& 30 &  0.84   \\
     Jiang and Conrath (IC ~\\ computed as Seco et al.\cite{5}) & IC (intrinsic) & 30  & 0.88  \\
  \hline 
     Resnik (IC ~\\computed as Zhou et al.\cite{7}) & IC (intrinsic)  & 30  &  0.85 \\
     Lin (IC computed ~\\ as Zhou et al.\cite{7}) &  IC (intrinsic)& 30 &  0.84 \\
     Jiang and Conarth ~\\ (IC computed as Zhou et al.\cite{7}) &  IC (intrinsic)& 30 & 0.83  \\
   \hline 
     Resnik (IC computed ~\\ as S\'{a}nchez et al.2011\cite{6}) & IC (intrinsic)  & 30  &  0.83  \\
      Lin (IC computed ~\\ as S\'{a}nchez et al.2011\cite{6}) &  IC (intrinsic)& 30 & 0.84\\
       Jiang and Conarth ~\\ (IC computed as ~\\ S\'{a}nchez et al.2011\cite{6}) &  IC (intrinsic)& 30 & 0.87 \\
   \hline 
      Resnik (IC computed ~\\ as David S\'{a}nchez 2012\cite{4}) & IC (intrinsic)  & 30  &  0.81 \\
      Lin (IC computed ~\\ as David S\'{a}nchez 2012\cite{4}) &  IC (intrinsic)& 30 & 0.84\\
      Jiang and Conarth ~\\ (IC computed as ~\\ David S\'{a}nchez 2012\cite{4}) &  IC (intrinsic)& 30 & 0.88 \\
    \hline 
      Montserrat Batet et al.\cite{47} (IC computed ~\\ as S\'{a}nchez et al.2011\cite{6} ) & IC (intrinsic)  & 30  &  0.86   \\
    \hline 
       Resnik (IC computed ~\\ as Meng et al.\cite{8} ) & IC (intrinsic)  & 30  &  0.86   \\
      Lin (IC computed ~\\ as Meng et al.\cite{8}) &  IC (intrinsic)& 30 & 0.86 \\
      Jiang and Conarth ~\\ (IC computed as Meng et al.\cite{8}) &  IC (intrinsic)& 30 & 0.83 \\
   \hline 
      Resnik (IC computed ~\\ as Qingbo et al.\cite{50} ) & IC (intrinsic)  & 30  &  0.83   \\
      Lin (IC computed ~\\ as Qingbo et al.\cite{50}) &  IC (intrinsic)& 30 & 0.84 \\
      Jiang and Conarth ~\\ (IC computed as Qingbo et al.\cite{50}) &  IC (intrinsic)& 30 & 0.82 \\
   \hline 
    Pirr\'{o} \cite{48} (IC computed ~\\ as Seco et al.\cite{5} ) & IC (intrinsic)  & 30  &  0.83   \\
   \hline 
   
     \rowcolor{lightgray}
     Resnik (IC computed & & &\\ 
     \rowcolor{lightgray} by our model) & IC (intrinsic)  & 30  & 0.86    \\
     \rowcolor{lightgray}
     Lin (IC computed & & &\\ 
     \rowcolor{lightgray} by our model) &  IC (intrinsic)& 30 & 0.86\\
    \rowcolor{lightgray}
    Jiang and Conarth & & &\\ 
    \rowcolor{lightgray}(IC computed by our model) &  IC (intrinsic)& 30 & 0.84 \\
    \rowcolor{lightgray}
    Pirr\'{o} & & &\\ 
    \rowcolor{lightgray}(IC computed by our model) &  IC (intrinsic)& 30 & 0.86 \\
    \hline    
     \rowcolor{lightgray}
     Our\_Sim (IC computed & & &\\ 
     \rowcolor{lightgray} by our model) & IC (intrinsic)  & 30  &  0.86    \\
      \rowcolor{lightgray}
    Our\_Sim & & &\\ 
    \rowcolor{lightgray}(IC computed by Meng et al. model) &  IC (intrinsic)& 30 & 0.87 \\
      [1ex] 
      \hline
\end{tabular}
}
\end{center}
\end{table}
\noindent Few authors \cite{4} have reported evaluation results for this benchmark, due to its recentness. 
Correlation values are shown in \emph{TABLE \ref{6correlation201}} for this new benchmark data set. 
From this \emph{TABLE \ref{6correlation201}} we can clearly 
say our semantic similarity measure based on our IC model gives a significant result over existing semantic similarity measures. 
It is also noticeable that Pirro's semantic similarity model with our IC model 
gives best correlation value (i.e. 0.71) among all the existing semantic similarity 
model evaluated over Wordism similarity goldstandard benchmark data set. $Fig.~\ref{plot6intrinsicvsintrinsic201data}$ 
shows the comparison of correlation coefficient values of our proposed semantic
similarity measure embedded with Meng et al. and our IC
model versus state of the art intrinsic IC based semantic similarity
measures. It also shows correlation coefficient values of some of the existing similarity measures embedded 
with our intrinsic IC model versus state of the art intrinsic
IC based semantic similarity measures based on goldstandard
data set.
$X$ axis shows different semantic similarity measures. $Y$ axis shows 
correlation coefficient value with the goldstandard data set.\\

\indent During our experiment, we also observe that correlation scores are inversely proportional to the size of data set. For 28 number data, 
we get upper value 0.90. When size of the data set becomes 30, we get upper limit of correlation 0.88.
For 201 noun pairs correlation reaches maximum 0.71 by our
proposed IC model with Pirr\'{o} et al. semantic similarity model. 

\begin{table}[htbp]
 \caption{Shows correlation coefficient values for different non-IC based semantic similarity measures,
 and correlation values
obtained by our framework.}
   \label{5nonicbasedmethods}
 \begin{center}
 \fontsize{10}{0} \scalebox{1}{
 \begin{tabular}{p{3.2cm} p{1.8cm} p{.8cm} p{1.1cm}} 
 \hline
  Similarity measures&Type & No. of&Correlation\\ [0.5ex] 
  (with IC model)&   &noun pairs&with M\&C\\
 \hline\hline
    \hline 
     Rada et al.\cite{10} & Edge-counting  & 28  & 0.59   \\ 
     Wu and Palmer\cite{11}  & Edge-counting  & 28  & 0.74   \\
    Leacock and Chodorow\cite{12}   & Edge-counting  & 28  & 0.74   \\
     Li et al.\cite{13}  & Edge-counting  & 28  & 0.82  \\
     Rodriguez and Egenhofer\cite{14}  & Feature-based & 28  & 0.71  \\
    Tversky\cite{15}   & Feature-based  & 28  & 0.73   \\
     Petrakis et al. \cite{16} & Feature-based  & 30  & 0.73   \\
      Aida Valls et al. \cite{46} & Feature-based  & 30  & 0.83   \\
     Bollegala et al.\cite{17}  & Distributional  & 30  & 0.83   \\
    Chen et al.\cite{18}   & Distributional  & 30  & 0.69   \\
    Sahami and Heilman \cite{19}  & Distributional  & 30  & 0.58   \\
    Gledson \cite{45}  & Distributional & 30  & 0.55   \\
     Danushka Bollegala \cite{44}  & WebSnippest\& & & \\ &Page-count based & 28  & 0.87   \\
    \hline
    \rowcolor{lightgray}
     Resnik (IC computed & & &\\ 
     \rowcolor{lightgray} by our model) & IC (intrinsic)  & 30  & 0.86    \\
     \rowcolor{lightgray}
     Lin (IC computed & & &\\ 
     \rowcolor{lightgray} by our model) &  IC (intrinsic)& 30 & 0.86\\
    \rowcolor{lightgray}
    Jiang and Conarth & & &\\ 
    \rowcolor{lightgray}(IC computed by our model) &  IC (intrinsic)& 30 & 0.84 \\
     \rowcolor{lightgray}
      \rowcolor{lightgray}
    Pirr\'{o} & & &\\ 
    \rowcolor{lightgray}(IC computed by our model) &  IC (intrinsic)& 30 & 0.86 \\
    \hline    
     \rowcolor{lightgray}
     Our\_Sim (IC computed & & &\\ 
     \rowcolor{lightgray} by our model) & IC (intrinsic)  & 30  &  0.86    \\
      \rowcolor{lightgray}
    Our\_Sim & & &\\ 
    \rowcolor{lightgray}(IC computed by Meng et al. model) &  IC (intrinsic)& 30 & 0.87 \\
      [1ex] 
      \hline
\end{tabular}
}
\end{center}
\end{table}

\section{Conclusion}
We propose a framework comprises a novel semantic similarity finding approach and an intrinsic IC calculation model.
Our framework uses hyponym-hypernym relationship between concepts.
This framework captures detailed structural aspects of
WordNet ontology. Experimental results show significance of using those aspects on deciding amount of information contained in each concepts more
accurately.
Proposed semantic similarity model shows its power for measuring semantic similarity between any two concepts of WordNet.
Our intrinsic IC calculation approach overcomes the issues mentioned in section 2 and 3 to become superior IC calculation model. Our similarity model 
along with our proposed IC model gives significant correlation with larger benchmark data set. It also gives better results than most of the state of the art 
measures. Even Pirr\'{o} et al. similarity model embedded with our IC calculator gives so far the best result when tested with bigger data set. 
\emph{TABLE~\ref{5nonicbasedmethods}} shows that proposed IC calculation model and our semantic similarity calculation approach outperforms 
all other different non-IC based state of art semantic similarity calculation models used so far.
Apart from our IC model, our proposed semantic similarity model also gives very high correlation when embedded with Meng et al. IC model.
Experimental results show our proposed IC calculation model or semantic similarity calculation model is compatible enough to produce significant 
semantic similarity scores when embedded 
with others semantic similarity calculator or IC calculator respectively.\\
\indent Apart from the mentioned achievements, we have the following goals to achieve as our future work.
\begin{itemize}
\item We use WordNet as an underlying ontology in current work. In future work we want to upgrade proposed 
approach for handling knowledge base like Probase \cite{49, 43}.
Because Probase is more than a traditional ontology or taxonomy.
It has a concept space (2.7 million categories) bigger than WordNet. Concepts of Probase are automatically 
acquired from web pages authored by millions of users and search logs. 
\item We will try to design a novel approach for finding semantic similarity measure between two concepts based on multiple ontologies.
\end{itemize}
\begin{table}[htbp]
  \caption{Shows correlation coefficient values for intrinsic IC-based state of the art semantic similarity measures based on
  different intrinsic IC calculators and correlation values
obtained by our proposed framework.}
   \label{6correlation201}
 \begin{center}
 \fontsize{10}{0} \scalebox{1}{
 \begin{tabular}{p{3.2cm} p{1.8cm} p{.9cm} p{1.1cm}} 
 \hline
  Similarity measures&Type & No. of&Correlation\\ [0.5ex] 
  (with IC model)&   &noun pairs&with Wordism\\
 \hline\hline
     Resnik (IC ~\\computed as Seco et al.\cite{5}) & IC (intrinsic) & 201 &  0.66 \\
     Lin (IC ~\\ computed as Seco et al.\cite{5}) &  IC (intrinsic)&  201  &   0.69  \\
     Jiang and Conrath (IC ~\\ computed as Seco et al.\cite{5}) & IC (intrinsic) &  201  & 0.67 \\
  \hline 
     Resnik (IC ~\\computed as Zhou et al.\cite{7}) & IC (intrinsic)  &  201   & 0.64 \\
     Lin (IC computed ~\\ as Zhou et al.\cite{7}) &  IC (intrinsic)&  201  &  0.64 \\
     Jiang and Conarth ~\\ (IC computed as Zhou et al.\cite{7}) &  IC (intrinsic)&  201  &  0.63  \\
   \hline 
     Resnik (IC computed ~\\ as S\'{a}nchez et al.2011\cite{6}) & IC (intrinsic)  &  201  &  0.66   \\
      Lin (IC computed ~\\ as S\'{a}nchez et al.2011\cite{6}) &  IC (intrinsic)&  201  &  0.66\\
       Jiang and Conarth ~\\ (IC computed as ~\\ S\'{a}nchez et al.2011\cite{6}) &  IC (intrinsic)&  201  & 0.66 \\
   \hline 
      Resnik (IC computed ~\\ as David S\'{a}nchez 2012\cite{4}) & IC (intrinsic)  &  201   &  0.67   \\
      Lin (IC computed ~\\ as David S\'{a}nchez 2012\cite{4}) &  IC (intrinsic)&  201  & 0.69\\
       Jiang and Conarth ~\\ (IC computed as ~\\ David S\'{a}nchez 2012\cite{4}) &  IC (intrinsic)&  201  & 0.67\\
   \hline 
      Montserrat Batet et al. \cite{47} (IC computed ~\\ as S\'{a}nchez et al.2011\cite{6} ) & IC (intrinsic)  & 201  &  0.68   \\
   \hline 
       Resnik (IC computed ~\\ as Meng et al.\cite{8} ) & IC (intrinsic)  &  201   &  0.67   \\
      Lin (IC computed ~\\ as Meng et al.\cite{8}) &  IC (intrinsic)&  201  & 0.68\\
      Jiang and Conarth ~\\ (IC computed as Meng et al.\cite{8}) &  IC (intrinsic)&  201  &0.66 \\
   \hline 
   Pirr\'{o} \cite{48} ~\\ (IC computed as Seco et al.\cite{8}) &  IC (intrinsic)&  201  & 0.70 \\
   \hline
       Resnik (IC computed ~\\ as Qingbo et al.\cite{50} ) & IC (intrinsic)  &  201   &  0.68   \\
      Lin (IC computed ~\\ as Qingbo et al.\cite{50}) &  IC (intrinsic)&  201  & 0.69\\
      Jiang and Conarth ~\\ (IC computed as Qingbo et al.\cite{50}) &  IC (intrinsic)&  201  &0.65 \\
   \hline 
   
   \rowcolor{lightgray}
     Resnik (IC computed & & &\\ 
     \rowcolor{lightgray} by our model) & IC (intrinsic)  &  201   &  0.68  \\
     \rowcolor{lightgray}
     Lin (IC computed & & &\\ 
     \rowcolor{lightgray} by our model) &  IC (intrinsic)&  201  & 0.68\\
    \rowcolor{lightgray}
    Jiang and Conarth & & &\\ 
    \rowcolor{lightgray}(IC computed by our model) &  IC (intrinsic)& 201  & 0.66 \\
     \rowcolor{lightgray}
    Pirr\'{o} & & &\\ 
    \rowcolor{lightgray}(IC computed by our model) &  IC (intrinsic)& 201  & 0.71\\
     \rowcolor{lightgray}
      \hline 
    Our\_{sim} & & &\\ 
    \rowcolor{lightgray}(IC computed by our model) &  IC (intrinsic)& 201  & 0.69\\
     \rowcolor{lightgray}
     Our\_{sim} & & &\\ 
    \rowcolor{lightgray}(IC computed by Meng et al.) &  IC (intrinsic)& 201  & 0.68 \\
      [1ex] 
      \hline
\end{tabular}
}
\end{center}
\end{table}

  \section*{Acknowledgment}
This work is supported by the Visvesvaraya PhD scheme, Sponsored by DeitY, MCIT, Govt. of India.

\ifCLASSOPTIONcaptionsoff
  \newpage
\fi

\onecolumn
\appendices
\section{}

\begin{table}[!ht]
 \caption{semantic similarity scores and correlation coefficient values obtained by our intrinsic IC calculation approach (\emph{eq.17}) when applied to Resnik's,
 Lin's, Jiang and Conarth's, Pirro's and Our's (\emph{eq.18}) similarity measures for each noun pairs of Wordism Similarity Goldstandard benchmark data set. }
\label{T7similarityscore201data}
  \begin{center}
 \fontsize{10}{0} \scalebox{1}{
 \begin{tabular}{||p{4cm} p{1.5cm} p{1.5cm} p{2cm} p{1.5cm} p{1.5cm} p{1.5cm}||} 
 \hline
 Noun pairs &Resnik(IC computed as eq.17) & Lin(IC computed as eq.17) & Jiang\&Conarth (IC computed as eq.17) & Pirr\'{o}(IC computed as eq.17)& Our Sim(IC computed as eq.17)& Our Sim(IC computed as Meng)\\
 \hline\hline
 1.tiger-cat & 0.827  &0.956  & 0.962 & 0.916 & 0.887  & 0.882 \\ 
 \hline
 2. tiger-tiger&  0.902 & 1.000 &1.000  &  1.000& 0.948 & 0.945 \\ 
 \hline
 3.plane-car   &0.460   &0.654  & 0.756 & 0.486& 0.540 & 0.521 \\ 
 \hline
 4.train-car  & 0.424  &0.625  &  0.746&  0.455& 0.505 & 0.481 \\ 
 \hline
 5.television-radio  &0.736   & 0.916 &0.932  & 0.845 & 0.816 &0.801  \\ 
 \hline
 6.media-radio  & 0.538  & 0.816 & 0.878 & 0.689 & 0.652 &  0.623\\ 
 \hline
  7. bread-butter& 0.379  & 0.706 & 0.842 &  0.545& 0.445 & 0.410 \\ 
\hline
 8.cucumber-potato  & 0.525  &  0.704& 0.797 & 0.543 & 0.579 & 0.564 \\ 
\hline
 9.doctor-nurse  & 0.492  & 0.810 & 0.884 & 0.681 &  0.613&  0.584\\ 
\hline
10.professor-doctor  & 0.414  & 0.640 & 0.767 &  0.470& 0.503 &  0.467\\ 
\hline
11.student-professor & 0.161  & 0.286 & 0.598 & 0.167 & 0.208 &  0.174\\ 
\hline
12.smart-stupid & 0.000  & 0.000 & 0.333 &0.000 & 0.000 &  0.000\\ 
\hline
13.wood-forest &0.564   & 1.000 & 1.000 & 1.000 & 0.721 &  0.680\\ 
\hline
14. money-cash 	& 0.534  & 0.777 &  0.846& 0.635 & 0.633 &  0.605\\ 
\hline
15. king-queen &  0.749  & 1.000 & 1.000 & 1.000 & 0.826 &  0.811\\ 
\hline
16. king-rook & 0.741 &0.920  & 0.935 & 0.852 & 0.826 &  0.811\\ 
\hline
17.bishop-rabbi   & 0.435  &0.686  & 0.801 & 0.522 & 0.533 &  0.498\\ 
\hline
18. fuck-sex   &0.483  &0.813  & 0.888 & 0.685 & 0.609 &  0.571\\ 
\hline
19.football-soccer  & 0.753   & 0.961 & 0.969 & 0.924 & 0.844 & 0.830 \\ 
\hline
20.football-basketball  & 0.643  &0.823  & 0.861 & 0.699 & 0.722 &  0.705\\ 
\hline
21. football-tennis& 0.574 & 0.760 & 0.819 & 0.614 & 0.654 &  0.631\\ 
\hline
22. Arafat-Jackson &   0.427 & 0.625 & 0.744 & 0.455 & 0.508 & 0.475 \\ 
\hline
23. physics-chemistry  & 0.534 &0.772  &  0.842& 0.629 & 0.631 & 0.616 \\ 
\hline
24. vodka-gin&  0.598  & 0.811 &0.861  & 0.683 &0.688  &  0.666\\ 
\hline
25. vodka-brandy  & 0.598  &0.826  & 0.874 & 0.704 & 0.694 & 0.672 \\ 
\hline
26. car-automobile & 0.704 & 1.000 & 1.000 &1.000  & 0.826 &  0.815\\ 
\hline
27. gem-jewel   &  0.701 &1.000  & 1.000 & 1.000 & 0.824 &  0.806\\ 
\hline
28. journey-voyage& 0.663  & 0.905 & 0.930 &0.826  & 0.766 & 0.753 \\ 
\hline
29. boy-lad  & 0.537  & 0.897 &  0.938& 0.813 & 0.673 &  0.633\\ 
\hline
30.coast-shore    & 0.491  & 0.935 & 0.966 & 0.879 & 0.645 & 0.587 \\ 
\hline
31.  asylum-madhouse & 0.775  & 0.975 & 0.980 & 0.952 & 0.864 & 0.849 \\ 
\hline
32. magician-wizard   & 0.564  & 1.000 & 1.000 & 1.000 & 0.721 &  0.680\\ 
\hline
33. midday-noon  & 0.782  & 1.000 &1.000  & 1.000 & 0.877 &  0.863\\ 
\hline
34. furnace-stove    & 0.199  &0.274  & 0.473 &  0.159& 0.231 & 0.207 \\ 
\hline
35.  food-fruit  & 0.128  & 0.190 & 0.649 & 0.105 & 0.153 & 0.103 \\ 
\hline
36. bird-cock   &  0.488 & 0.753 &  0.840& 0.604 &0.598  &  0.587\\ 
\hline
37.  bird-crane  & 0.488  & 0.721 & 0.811 & 0.564 & 0.589 &  0.578\\ 
\hline
38.  food-rooster&  0.048 &  0.082 &0.457  & 0.042 & 0.063 &  0.000\\ 
\hline
39.money-dollar& 0.534  & 0.727 &  0.800& 0.572 &0.616  &  0.587\\ 
\hline
40.money-currency &  0.534 & 0.866 & 0.917 &0.764 & 0.662 & 0.634 \\ 
\hline
41.tiger-jaguar &  0.827 & 0.908 &0.916  & 0.832 &0.866  &  0.860\\ 
\hline
42.tiger-feline& 0.734  & 0.897 & 0.916 & 0.813 & 0.809 &  0.802\\ 
\hline
43.tiger-carnivore & 0.584  & 0.786 &0.841  & 0.647& 0.676 &  0.668\\ 
\hline
44.tiger-mammal& 0.470  &0.684  & 0.783 &  0.520& 0.566 &  0.554\\ 
\hline
45.tiger-animal &  0.322 & 0.526 & 0.751 & 0.357 & 0.413 &  0.392\\ 
\hline
46.tiger-organism&  0.185 & 0.507 & 0.819 & 0.340 &0.277  &  0.264\\ 
\hline
47.tiger-fauna  &  0.322 & 0.526 & 0.751 & 0.357 & 0.413 & 0.392 \\ 
\hline
48.psychology-psychiatry& 0.473  & 0.622 &  0.712& 0.451 & 0.539 & 0.519 \\ 
\hline
49.psychology-science  & 0.473  &  0.824&  0.898& 0.700 & 0.604 &  0.584\\ 
\hline
50.psychology-discipline & 0.409  & 0.754 & 0.866 & 0.605 & 0.534 &  0.510\\ 
\hline
51.planet-star  & 0.468  & 0.821 & 0.897 &  0.696& 0.596 &  0.558\\ 
\hline
52.planet-moon& 0.468  &0.724  &0.821  &  0.568& 0.569 &  0.532\\ 
\hline
53.planet-sun& 0.468  & 0.720 & 0.818 & 0.563 & 0.568 &  0.531\\ 
\hline
54.precedent-example & 0.554  &0.912  &0.946  & 0.839 & 0.690 & 0.650 \\ 
\hline
55.precedent-antecedent & 0.254  & 0.401 &0.620  & 0.251 & 0.311 & 0.259 \\ 
\hline
 \hline
\end{tabular}
}
\end{center}
\end{table} 

\begin{table*}[!ht]
 \caption{semantic similarity scores and correlation coefficient values obtained by our intrinsic IC calculation approach (\emph{eq.17}) when applied to Resnik's,
 Lin's, Jiang and Conarth's, Pirro's and Our's (\emph{eq.18}) semantic similarity measures for each noun pairs of Wordism Similarity Goldstandard benchmark data set. }
\label{T8similarityscore201data}
  \begin{center}
 \fontsize{10}{0} \scalebox{1}{
 \begin{tabular}{||p{4cm} p{1.5cm} p{1.5cm} p{2cm} p{1.5cm} p{1.5cm} p{1.5cm}||} 
 \hline
 Noun pairs &Resnik(IC computed as eq.17) & Lin(IC computed as eq.17) & Jiang\&Conarth (IC computed as eq.17) & Pirr\'{o}(IC computed as eq.17)& Our Sim(IC computed as eq.17)& Our Sim(IC computed as Meng)\\
 \hline\hline
56.cup-tableware  &  0.555 &  0.919& 0.951 & 0.851 & 0.693 &  0.672\\ 
\hline
57.cup-artifact & 0.199  & 0.467 &0.773  & 0.305 &0.286  &  0.259\\ 
\hline
58.cup-object  & 0.155  & 0.242 & 0.718 & 0.137 & 0.183 & 0.121 \\ 
\hline
59.cup-entity  &0.000   & 0.000 & 0.695 & 0.000 &  0.000& 0.000 \\ 
\hline
60.jaguar-cat  & 0.827  & 0.947 & 0.954 & 0.899 & 0.884 &  0.878\\ 
\hline
61.jaguar-car & 0.121  & 0.154 &0.331  & 0.083 &0.136  &  0.116\\ 
\hline
62.mile-kilometer&  0.471 & 0.690 &  0.788&0.526  &0.560  &  0.524\\ 
\hline
63.skin-eye & 0.284  & 0.430 & 0.624 &0.274  & 0.342 &  0.306\\ 
\hline
64.Japanese-American& 0.335  &  0.610&  0.785& 0.439 & 0.434 &  0.391\\ 
\hline
65. century-year & 0.320  &0.564  & 0.752 & 0.393 & 0.409 &  0.370\\ 
\hline
66.announcement-news  & 0.254  & 0.467 &0.709  & 0.304  &0.329  & 0.274 \\ 
\hline
67.doctor-personnel &  0.050 &0.081  & 0.469 & 0.042&0.062  & 0.000 \\ 
\hline
68.Harvard-Yale   & 0.641  & 0.859 & 0.895 & 0.753 & 0.734 &  0.711\\ 
\hline
69.hospital-infrastructure & 0.050  & 0.073 &0.366  & 0.038 &0.059  & 0.000 \\ 
\hline
70.life-death  & 0.495  & 0.776 & 0.857 & 0.634 & 0.605 & 0.567 \\ 
\hline
71.travel-activity & 0.220  &  0.506&0.785  & 0.339 & 0.310 &  0.278\\ 
\hline
72.type-kind    & 0.646  & 0.916 & 0.941 & 0.846 & 0.759 &  0.741\\ 
\hline
73.street-place &  0.211 & 0.333 &0.576  & 0.200 &0.258  &  0.216\\ 
\hline
74.street-avenue & 0.659  & 0.922 &  0.944& 0.856  & 0.769 &  0.752\\ 
\hline
75. street-block  &0.199   & 0.343 & 0.618 & 0.207 & 0.252 &  0.229\\ 
\hline
76.cell-phone & 0.704  & 0.927 &0.945  &  0.865 &  0.801&  0.782\\ 
\hline
77.dividend-payment&  0.542 & 0.818 & 0.880 &0.693 & 0.655 &  0.634\\ 
\hline
78.calculation-computation    & 0.676  & 1.000 & 1.000 & 1.000  &0.807  & 0.789 \\ 
\hline
79.profit-loss&  0.486 & 0.728 & 0.818 &0.572 & 0.583 &  0.543\\ 
\hline
80. dollar-yen  & 0.388  & 0.606 & 0.748 & 0.435 & 0.474 & 0.443 \\ 
\hline
81.dollar-buck  & 0.782  & 1.000& 1.000 & 1.000 &  0.877&  0.863\\ 
\hline
82.phone-equipment & 0.425  &0.753  &0.860  & 0.604 & 0.548 &  0.523\\ 
\hline
83.liquid-water & 0.463  & 0.928 &0.964  & 0.867 & 0.619 & 0.562 \\ 
\hline
84.marathon-sprint &  0.220 & 0.284 &0.446  & 0.165 & 0.248 & 0.222 \\ 
\hline
85.seafood-food & 0.304  &0.801  & 0.924 & 0.668 & 0.442 &  0.399\\ 
\hline
86.seafood-lobster & 0.455  & 0.801 & 0.887 & 0.668 &0.583  & 0.546 \\ 
\hline
87.lobster-food   & 0.304  & 0.617 & 0.811 & 0.447&0.414  &  0.373\\ 
\hline
88.lobster-wine  &0.146   &  0.234&0.521  &  0.132 & 0.180 & 0.119 \\ 
\hline
89.championship-tournament & 0.456  & 0.705 &0.809  & 0.544 &0.554  &  0.520\\ 
\hline
90.man-woman    & 0.328  &  0.724& 0.875 &0.568 &0.451  & 0.407 \\ 
\hline
91.man-governor   & 0.262  & 0.360 &0.648  &0.219  &0.304  &  0.283\\ 
\hline
92.murder-manslaughter&0.721   &  0.900& 0.920 & 0.819 & 0.801 & 0.789 \\ 
\hline
93.opera-performance&  0.050 &0.082  & 0.447 & 0.042 &0.062  &  0.000\\ 
\hline
94.Mexico-Brazil & 0.500  & 0.640 & 0.718 & 0.470 &  0.561& 0.542 \\ 
\hline
95.glass-metal&  0.146 & 0.319 &0.687 & 0.190 & 0.201 &  0.135\\ 
\hline
96.aluminum-metal&  0.445 & 0.816 & 0.899 & 0.690 & 0.571 & 0.543 \\ 
\hline
97.rock-jazz   & 0.507  & 0.814 & 0.884 &  0.686& 0.625 &  0.586\\ 
\hline
98.museum-theater &  0.199 &  0.296& 0.527 &0.174  & 0.238 & 0.215 \\ 
\hline
99.shower-thunderstorm  & 0.483  & 0.622 &  0.706& 0.451 &0.544  &  0.517\\ 
\hline
100.monk-oracle   & 0.161  & 0.248 & 0.511 & 0.141 & 0.194 &  0.163\\ 
\hline
101.cup-food & 0.291  & 0.549 &  0.761& 0.379 & 0.390 &  0.350\\ 
\hline
102.journal-association &0.050   & 0.094 & 0.519 & 0.049 & 0.065 & 0.000 \\ 
\hline
103.street-children & 0.121  & 0.206 &  0.531& 0.115 &0.153  &  0.132\\ 
\hline
104.car-flight     & 0.199  &  0.291& 0.515 & 0.170 & 0.236 &  0.213\\ 
\hline
105.space-chemistry  & 0.050  & 0.106 & 0.575 & 0.056 &  0.068&  0.000\\ 
\hline
106.situation-conclusion &0.220   & 0.402 & 0.674 &0.252  &  0.285& 0.255 \\ 
\hline
107.word-similarity  &  0.275 &0.343  & 0.560 & 0.207 & 0.305 & 0.271 \\ 
\hline
108.peace-plan  & 0.050  & 0.092 & 0.506 & 0.048 &  0.065&  0.000\\ 
\hline
109.consumer-energy & 0.048  & 0.097 & 0.551 & 0.051 & 0.064 & 0.000 \\ 
\hline
110.ministry-culture & 0.235  & 0.361 & 0.584 & 0.220 & 0.286 & 0.237 \\ 
\hline
 \hline
\end{tabular}
}
\end{center}
\end{table*} 

\begin{table*}[!ht]
 \caption{semantic similarity scores and correlation coefficient values obtained by our intrinsic IC calculation approach (\emph{eq.17}) when applied to Resnik's,
 Lin's, Jiang and Conarth's, Pirro's and Our's (\emph{eq. 18}) semantic similarity measures for each noun pairs of Wordism Similarity Goldstandard benchmark data set. }
\label{T9similarityscore201data}
  \begin{center}
  \fontsize{10}{0} \scalebox{1}{
 \begin{tabular}{||p{4cm} p{1.5cm} p{1.5cm} p{2cm} p{1.5cm} p{1.5cm} p{1.5cm}||} 
 \hline
Noun pairs & Resnik(IC computed as eq.17) & Lin(IC computed as eq.17) & Jiang\&Conarth (IC computed as eq.17) & Pirr\'{o}(IC computed as eq.17)& Our Sim(IC computed as eq.17)& Our Sim(IC computed as Meng)\\
 \hline\hline
111.smart-student  & 0.000  & 0.000& 0.422 & 0.000 & 0.000 &  0.000\\ 
\hline
112.investigation-effort & 0.424  &0.739  & 0.850 & 0.586  & 0.539 &  0.514\\ 
\hline
113.image-surface& 0.275  & 0.384 & 0.662 &  0.237& 0.321 & 0.288 \\ 
\hline
114.life-term  &0.553   &0.878  & 0.923 &0.783 & 0.680 &  0.638\\ 
\hline
115.start-match  & 0.181  & 0.307 &0.590  & 0.181 & 0.228 & 0.190 \\ 
\hline
116.computer-news    & 0.000  & 0.000 &0.442  & 0.000 &  0.000& 0.000 \\ 
\hline
117.board-recommendation&0.050   & 0.078 & 0.413 & 0.040 & 0.061 &  0.000\\ 
\hline
118.lad-brother & 0.161  & 0.268 & 0.561 & 0.155 & 0.201 & 0.170 \\ 
\hline
119.observation-architecture    & 0.288  & 0.399 & 0.565 &  0.249 & 0.334 & 0.312 \\ 
\hline
120.coast-hill  &  0.301 &0.553  & 0.756 & 0.382 & 0.390 &  0.325\\ 
\hline
121.deployment-departure &0.220   & 0.361 & 0.610 &  0.220&0.274  &  0.248\\ 
\hline
122.benchmark-index & 0.409  &0.648  &0.778  & 0.480 & 0.501 &  0.453\\ 
\hline
123.attempt-peace  &0.050   &0.092  &0.508  &  0.048& 0.065 & 0.000 \\ 
\hline
124.consumer-confidence& 0.000  & 0.000 &0.428  & 0.000 & 0.000 &  0.000\\ 
\hline
125.start-year & 0.176  & 0.332 & 0.647 &0.199 & 0.230 &  0.154\\ 
\hline
126.focus-life    &  0.211 & 0.318 & 0.547 & 0.189 & 0.254 &  0.211\\ 
\hline
127.development-issue &  0.307 &0.477 & 0.740 & 0.313 & 0.373 & 0.336 \\ 
\hline
128.theater-history & 0.128  & 0.214 &0.528  &0.119  & 0.160 & 0.109 \\ 
\hline
129.situation-isolation &0.220   & 0.435 & 0.725 & 0.278 & 0.285 & 0.250\\ 
\hline
130.profit-warning &  0.050 & 0.077 & 0.403 & 0.040 & 0.061 & 0.000 \\ 
\hline
131.media-trading  &0.220   &0.333  &  0.559& 0.200 & 0.265 &  0.239\\ 
\hline
132.chance-credibility  & 0.140  & 0.258 & 0.597 & 0.148 &  0.181& 0.122 \\ 
\hline
133.precedent-information& 0.332  & 0.668 & 0.835 & 0.502  & 0.449 & 0.405 \\ 
\hline
134.architecture-century& 0.050  & 0.081 &  0.430& 0.042 & 0.062 & 0.000 \\ 
\hline
135.population-development& 0.470  & 0.592 &0.675  &0.420 & 0.524 & 0.505 \\ 
\hline
136.peace-atmosphere &0.273   & 0.430 & 0.644 & 0.274 & 0.335 &  0.302\\ 
\hline
137.morality-marriage  &0.140   & 0.280 &  0.641& 0.163 &0.188  &  0.126\\ 
\hline
138.minority-peace   & 0.211  &0.344  &0.596  & 0.207 & 0.262 & 0.218 \\ 
\hline
139.atmosphere-landscape &0.215   &0.311  &0.524  & 0.184 & 0.255 &  0.212\\ 
\hline
140.report-gain  & 0.050  &  0.087& 0.477 & 0.045 & 0.064 &  0.000\\ 
\hline
141.music-project   & 0.288  & 0.501 &0.713  & 0.334 & 0.366 & 0.342 \\ 
\hline
142.seven-series     & 0.153  &0.231  & 0.489 & 0.130 & 0.185 &  0.124\\ 
\hline
143.experience-music & 0.207  &0.327  &  0.627& 0.195 &0.249  & 0.206 \\ 
\hline
144.school-center& 0.449  & 0.670 & 0.778 & 0.503 & 0.537 &  0.514\\ 
\hline
145.five-month& 0.176  & 0.296 & 0.582 & 0.174 & 0.222 &  0.146\\ 
\hline
146.announcement-production&  0.155 & 0.272 & 0.585 & 0.157 & 0.198 & 0.132 \\ 
\hline
147.morality-importance   & 0.242  &0.499  & 0.756 & 0.332 & 0.328 & 0.273 \\ 
\hline
148.money-operation  & 0.050  & 0.109 &0.588  &0.057  & 0.069 &  0.000\\ 
\hline
149.delay-news   & 0.181  & 0.280 &  0.533& 0.162 & 0.220 & 0.183 \\ 
\hline
150.governor-interview & 0.000  & 0.000 & 0.344 & 0.000 &0.000  &  0.000\\ 
\hline
151.practice-institution& 0.490  & 0.858 &  0.918& 0.751 &0.626  &  0.568\\ 
\hline
152.century-nation& 0.050  & 0.092 & 0.504 & 0.048  & 0.065 &  0.000\\ 
\hline
153.coast-forest &  0.090 &0.169  & 0.559 &0.092 &0.117  &  0.083\\ 
\hline
154.shore-woodland  & 0.090  & 0.181 & 0.593 & 0.099 & 0.120 &  0.085\\ 
\hline
155.drink-car   & 0.048  &0.090  & 0.515 &  0.047&  0.063& 0.000 \\ 
\hline
156.president-medal &  0.050 &0.072  & 0.387 & 0.038 & 0.059 &  0.000\\ 
\hline
157.prejudice-recognition   & 0.390  & 0.591 & 0.729 & 0.419 & 0.470 & 0.421 \\ 
\hline
158.viewer-serial & 0.199  & 0.261 & 0.465 &  0.150& 0.226 &  0.203\\ 
\hline
159.peace-insurance    &  0.630 & 0.868 & 0.904 & 0.767 & 0.730 & 0.697 \\ 
\hline
160.Mars-water &  0.121 & 0.173 & 0.596 & 0.094 &0.143  & 0.122 \\ 
\hline
161.media-gain& 0.153  & 0.295 & 0.632 & 0.173 & 0.202 &  0.135\\ 
\hline
162.precedent-cognition & 0.207  &0.477  & 0.773 & 0.313 & 0.296 & 0.247 \\ 
\hline
163.announcement-effort&  0.050 & 0.094 &  0.516&  0.049& 0.065 & 0.000 \\ 
\hline
164.line-insurance  &  0.370 & 0.623 & 0.776 & 0.453 & 0.464 & 0.417 \\ 
\hline
165.crane-implement & 0.462  & 0.463 & 0.696 & 0.302 &  0.339& 0.316 \\ 
\hline
 \hline
\end{tabular}
}
\end{center}
\end{table*}

\begin{table*}[!ht]
 \caption{semantic similarity scores and correlation coefficient values obtained by our intrinsic IC calculation approach (\emph{eq.17}) when applied to Resnik's,
 Lin's, Jiang and Conarth's, Pirro's and Our's (\emph{eq. 18}) semantic similarity measures for each noun pairs of Wordism Similarity Goldstandard benchmark data set. }
\label{T9similarityscore201data}
  \begin{center}
  \fontsize{10}{0} \scalebox{1}{
 \begin{tabular}{||p{4cm} p{1.5cm} p{1.5cm} p{2cm} p{1.5cm} p{1.5cm} p{1.5cm}||} 
 \hline
 Noun pairs &Resnik(IC computed as eq.17) & Lin(IC computed as eq.17) & Jiang\&Conarth (IC computed as eq.17) & Pirr\'{o}(IC computed as eq.17)& Our Sim(IC computed as eq.17)& Our Sim(IC computed as Meng)\\
 \hline\hline
166.drink-mother   &0.227   & 0.427 & 0.694 & 0.271 & 0.299 & 0.248 \\ 
\hline
167.opera-industry & 0.050  & 0.081 & 0.433 & 0.042 & 0.062 & 0.000 \\ 
\hline
168.volunteer-motto& 0.000  & 0.000 & 0.386 & 0.000  & 0.000 &  0.000\\ 
\hline
169.listing-proximity&  0.128 &0.170  & 0.422 & 0.093& 0.146 & 0.097 \\ 
\hline
170.precedent-collection &  0.292 & 0.649 & 0.842 & 0.481 & 0.408 &  0.343\\ 
\hline
171.cup-article &0.470   & 0.837 & 0.908 & 0.721 & 0.605 & 0.566 \\ 
\hline
172.sign-recess & 0.326  & 0.488 & 0.658 &  0.323& 0.391 &  0.365\\ 
\hline
173.problem-airport& 0.000  & 0.000 & 0.350 & 0.000 & 0.000 &  0.000\\ 
\hline
174.reason-hypertension& 0.273  & 0.374 & 0.541 & 0.230 & 0.316 &  0.282\\ 
\hline
175.direction-combination  &  0.288 & 0.401 & 0.616 & 0.251 &  0.336& 0.313 \\ 
\hline
176.Wednesday-news& 0.050  & 0.078 & 0.407 & 0.040 & 0.061 &  0.000\\ 
\hline
177.glass-magician& 0.123  & 0.202 & 0.530 & 0.112 &  0.153& 0.131 \\ 
\hline
178.cemetery-woodland& 0.090  & 0.145 &0.469  & 0.078 & 0.111 &  0.078\\ 
\hline
179.possibility-girl & 0.000  &0.000  & 0.465 & 0.000 & 0.000 & 0.000 \\ 
\hline
180.cup-substance &  0.227 &0.456  & 0.729 &0.296  &  0.314& 0.261 \\ 
\hline
181.forest-graveyard    &0.090   & 0.145 & 0.469 & 0.078 & 0.111 & 0.078 \\ 
\hline
182.stock-egg  &0.291   & 0.421 & 0.600 &  0.267& 0.344 &  0.309\\ 
\hline
183.month-hotel &0.000   & 0.000 & 0.438 & 0.000 & 0.000 & 0.000 \\ 
\hline
184.energy-secretary &  0.048 & 0.087 & 0.495 & 0.045 & 0.062 & 0.000 \\ 
\hline
185.precedent-group &0.136   & 0.365 & 0.763 & 0.223 & 0.204 & 0.138 \\ 
\hline
186.production-hike& 0.220  & 0.324 & 0.540 & 0.193 &0.263  &  0.238\\ 
\hline
187.stock-phone &  0.326 & 0.450 &0.602  & 0.290 & 0.379 &  0.362\\ 
\hline
188.holy-sex & 0.048  &0.084  & 0.476 & 0.044 & 0.061 & 0.000 \\ 
\hline
189.stock-CD &  0.326 & 0.461 & 0.618 &  0.300& 0.383 & 0.366 \\ 
\hline
190.drink-ear &  0.175 & 0.281 & 0.582 &  0.164&  0.216& 0.144 \\ 
\hline
191.delay-racism  & 0.220  &0.340  & 0.573 &  0.205&  0.267&  0.241\\ 
\hline
192.stock-life & 0.211  & 0.318 &  0.547& 0.189 & 0.254 &  0.211\\ 
\hline
193.stock-jaguar    & 0.492  & 0.558& 0.611 &0.387  & 0.524 &  0.515\\ 
\hline
194.monk-slave   & 0.161  & 0.291 &0.607  & 0.170 & 0.208 & 0.176 \\ 
\hline
195.lad-wizard &  0.161 & 0.277 & 0.579 &  0.160& 0.203 &  0.172\\ 
\hline
196.sugar-approach& 0.050  &0.085  &  0.480& 0.044 & 0.063 & 0.000 \\ 
\hline
197.rooster-voyage & 0.000  & 0.000 & 0.153 & 0.000 & 0.000 &  0.000\\ 
\hline
198.noon-string   & 0.050  & 0.077 & 0.396 & 0.040 &0.061  &  0.000\\ 
\hline
199.chord-smile  & 0.155  &0.227  &  0.471& 0.128 &0.184  &  0.122\\ 
\hline
200.professor-cucumber & 0.185  & 0.238 & 0.405 & 0.135 & 0.209 &  0.196\\ 
\hline
201.king-cabbages &  0.185 &  0.290& 0.545  &  0.169& 0.228 & 0.216 \\ 
\rowcolor{lightgray}
CORRELATION: & 0.68
 & 0.68  &0.66 &  0.71 &0.69 & 0.68
 \\
 \hline
\end{tabular}
}
\end{center}
\end{table*}

\vfill


\begin{thebibliography}{1}
  
\bibitem{1} P. Resnik,\textquotedblleft Using Information Content to Evaluate Semantic Similarity in a
Taxonomy,\textquotedblright~\emph{Proceedings of the 14th International Joint Conference on Artificial Intelligence}, Montreal, Quebec, Canada, Volume 1, pp. 448-453, 1995.   
  
\bibitem{2} J. J. Jiang and D. W. Conrath,\textquotedblleft Semantic Similarity Based on Corpus Statistics and Lexical
Taxonomy,\textquotedblright~\emph{Proceedings of International Conference Research on Computational Linguistics (ROCLING X)}, Taiwan, pp. 19-33, 1997.

\bibitem{3} D. Lin,\textquotedblleft An Information-Theoretic Definition of Similarity,\textquotedblright~ 
\emph{In proceedings of the Fifteenth International Conference on Machine Learning}, Madison, Wisconsin, USA, pp. 296-304, 1998.

\bibitem{4} D. S\'{a}nchez and M. Batet,\textquotedblleft A New Model to Compute the Information Content of Concepts from Taxonomic Knowledge,\textquotedblright~ 
\emph{International Journal on Semantic Web \& Information Systems archive}, Volume 8 Issue 2, pp. 34-50, 2012.

\bibitem{5} N. Seco, T. Veale, and J. Hayes,\textquotedblleft An Intrinsic Information Content Metric for Semantic Similarity in WordNet,\textquotedblright~
\emph{In proceedings of the 16th European Conference on Artificial Intelligence - ECAI}, Valencia, Spain, pp. 1089-1090, 2004.

\bibitem{6} D. S\'{a}nchez, M. Batet, and D. Isern, \textquotedblleft Ontology Based Information Content Computation,\textquotedblright~\emph{  
Journal on Knowledge-Based Systems}, Volume 24 Issue 2, pp. 297-303, 2011.

\bibitem{7} Z. Zhou, Y. Wang, and J. Gu,\textquotedblleft A New Model of Information Content for Semantic Similarity in WordNet,\textquotedblright~
\emph{International Conference on Future Generation Communication and Networking Symposia}, IEEE Computer Society, pp. 85-89, 2008.

\bibitem{8} L. Meng, J. Gu, and Z. Zhou,\textquotedblleft A New Model of Information Content Based on Concept’s Topology
for Measuring Semantic Similarity in WordNet,\textquotedblright~\emph{International Journal of Grid and Distributed Computing}, Vol. 5, No. 3, pp. 81-94, 2012.

\bibitem{9} G. Miller and W.G. Charles,\textquotedblleft Contextual Correlates of Semantic
Similarity,\textquotedblright~\emph{Journal of Language and Cognitive Processes}, Vol. 6, pp. 1-28, 1991.

\bibitem{10} R. Rada, H. Mili, E. Bichnell, and M. Blettner,\textquotedblleft Development and Application of a
Metric on Semantic Nets,\textquotedblright~\emph{IEEE Transaction on Systems, Man and Cybernetics}, Vol. 19, Issue 1, pp. 17-30, 1989.

\bibitem{11} Z. Wu, and M. Palmer,\textquotedblleft Verb Semantics and Lexical Selection,\textquotedblright~\emph{In Proceeding of 32nd
annual Meeting of the Association for Computational Linguistics}, Association
for Computational Linguistics, Las Cruces, New Mexico, pp. 133-138, 1994.

\bibitem{12} C. Leacock, M. Chodorow,\textquotedblleft Combining Local Context and WordNet Similarity for
Word Sense Identification,\textquotedblright~\emph{WordNet: An Electronic Lexical Database}, MIT Press, pp. 265-283, 1998.

\bibitem{13} Y. Li, Z. Bandar, and D. McLean,\textquotedblleft An Approach for Measuring Semantic Similarity
between Words using Multiple Information Sources,\textquotedblright~\emph{IEEE Transaction on Knowledge and Data Engineering}, Vol. 15, Issue 4, pp. 871-882, 2003.

\bibitem{14} M. A. Rodriguez, and M. J. Egenhofer,\textquotedblleft Determining Semantic Similarity among Entity
Classes from Different Ontologies,\textquotedblright~\emph{IEEE Transaction on Knowledge and Data Engineering}, Vol. 15, Issue 2, pp. 442-456, 2003.

\bibitem{15} A. Tversky,\textquotedblleft Features of Similarity,\textquotedblright~\emph{Psychological Review}, Vol. 84, Issue 4, pp. 327-352, 1977.

\bibitem{16} E. G. M. Petrakis, G. Varelas, A. Hliaoutakis, and P. Raftopoulou,\textquotedblleft X-similarity:
Computing Semantic Similarity between Concepts from Different Ontologies,\textquotedblright~
\emph{Journal of Digital Information Management}, Vol. 4, pp. 233-237, 2006.

\bibitem{17} D. Bollegala, Y. Matsuo, and M. Ishizuka,\textquotedblleft Measuring Semantic Similarity between
Words Using Web Search Engines,\textquotedblright~\emph{In Proceeding of 16th international conference on
World Wide Web}, ACM Press, Banff, Alberta, Canada, pp. 757-766, 2007.

\bibitem{18} H. H. Chen, M. S. Lin, and Y. C. Wei,\textquotedblleft Novel Association Measures Using Web Search
with Double Checking,\textquotedblright~\emph{In Proceeding of 21st International Conference on
Computational Linguistics and the 44th annual meeting of the Association
for Computational Linguistics}, Sydney, Australia, pp. 1009-1016, 2006.

\bibitem{19} M. Sahami, and T. D. Heilman,\textquotedblleft A Web-based Kernel Function for Measuring the
Similarity of Short Text Snippets,\textquotedblright~\emph{In Proceeding of 15th International World Wide
Web Conference}, ACM Press, Edinburgh, Scotland, pp. 377-386, 2006.

\bibitem{20} http://en.wikipedia.org/wiki/Correlation\_coefficient.

\bibitem{21} https://wordnet.princeton.edu.

\bibitem{22} http://www.helsinki.fi/varieng/CoRD/corpora/BROWN/.

 \bibitem{23} S. Patwardhan, S. Banerjee, and T. Pedersen,\textquotedblleft Using Measures of Semantic
 Relatedness for Word Sense Disambiguation,\textquotedblright~ \emph{In Proceeding of 4th International
 Conference on Computational Linguistics and Intelligent Text Processing and
 Computational Linguistics}, CICLing 2003, Springer Berlin/Heidelberg, Mexico
 City, Mexico, pp. 241-257, 2003.
 
 \bibitem{24} A. Budanitsky, and G. Hirst,\textquotedblleft Semantic Distance in WordNet: An Experimental,
 Application-oriented Evaluation of Five Measures,\textquotedblright~\emph{In Proceeding of Workshop on
 WordNet and Other Lexical Resources}, Second meeting of the North American
 Chapter of the Association for Computational Linguistics. Pittsburgh, USA, pp. 10-15, 2001.
 
 \bibitem{25} J. R. Curran,\textquotedblleft Ensemble Methods for Automatic Thesaurus Extraction,\textquotedblright~\emph{In Proceeding of
 Empirical Methods in Natural Language Processing}, EMNLP 2002. Association
 for Computational Linguistics, Philadelphia, PA, USA, pp. 222-229, 2002.
 
 \bibitem{26} J. Atkinson, A. Ferreira, and E. Aravena,\textquotedblleft Discovering Implicit Intention-level
 knowledge from Natural-language Texts,\textquotedblright~\emph{Knowledge Based System}, Vol. 22, Issue 7, pp. 502-508, 2009.
 
 \bibitem{27} M. Stevenson, and M. A. Greenwood,\textquotedblleft A Semantic Approach to IE Pattern Induction,\textquotedblright~
 \emph{In Proceeding of 43rd Annual Meeting on Association for Computational Linguistics, COLING-ACL 2005}, Ann Arbor,
 Michigan, USA, pp. 379-386, 2005.
 
 \bibitem{28} D. S\'{a}nchez, D. Isern, and M. Millan,\textquotedblleft Content Annotation for the Semantic Web: An
 Automatic Web-based Approach,\textquotedblright~\emph{Knowledge and Information Systems}, Vol. 27, Issue 3, pp. 393-418, 2011. 
 
 \bibitem{29} M. Gaeta, F. Orciuoli, and P. Ritrovato,\textquotedblleft Advanced Ontology Management System for
 Personalised E-Learning,\textquotedblright~\emph{Knowledge Based System}, Vol. 22, Issue 4, pp. 292-301, 2009.
 
 \bibitem{30} A. Formica,\textquotedblleft Concept Similarity in Formal Concept Analysis: An Information
 Content Approach,\textquotedblright~\emph{Knowledge Based System}, Vol. 21, Issue 1, pp. 80-87, 2008.
 
 \bibitem{31} R. Nayak, and W. Iryadi, \textquotedblleft XML Schema Clustering with Semantic and Hierarchical
 Similarity Measures,\textquotedblright~\emph{Knowledge Based System}, Vol. 20, Issue 4, pp. 336-349, 2007.
 
 \bibitem{32} A. G. Tapeh, and M. Rahgozar,\textquotedblleft A knowledge-based Question Answering System for
 B2C e-Commerce,\textquotedblright~\emph{Knowledge Based System}, Vol. 21, Issue 8, pp. 946-950, 2008.
 
 \bibitem{33} Y. B. Fernandez, J. J. P. Arias, A. G. Solla, M. R. Cabrer, M. L. Nores, J. G. Duque, A. F. Vilas, R. P. D. Redondo, 
 and J. B. Munoz,\textquotedblleft A flexible Semantic Inference Methodology to Reason about User
 Preferences in knowledge-based Recommender Systems,\textquotedblright~ \emph{ Knowledge Based System}, Vol. 21, Issue 4, pp. 305-320, 2008.
 
 \bibitem{34} J. Debenham, and C. Sierra,\textquotedblleft Merging Intelligent Agency and the Semantic Web,\textquotedblright~ \emph{Knowledge Based System}, Vol. 21, Issue 3, pp. 184-191, 2008.
 
 \bibitem{35} K. A. Spackman, \textquotedblleft SNOMED CT Milestones: Endorsements are Added to Already-impressive Standards Credentials,\textquotedblright~ 
 \emph{In Healthcare Informatics: the Business Magazine for Information and Communication Systems}, Vol. 21, pp. 54-56, 2004.
 
 \bibitem{36} S. J. Nelson, D. Johnston, and B. L. Humphreys, \textquotedblleft Relationships in Medical Subject Headings, \textquotedblright~
 \emph{In Relationships in the Organization of Knowledge, K.A. Publishers}, pp. 171-184, 2001.
 
 \bibitem{37} http://en.wikipedia.org/wiki/Semantic\_similarity\#cite\_note-20.

\bibitem{38} S. Harispe, S. Ranwez, S. Janaqi, and J. Montmain,\textquotedblleft Semantic Measures for the Comparison of
Units of Language, Concepts or Instances from Text and Knowledge Representation Analysis,\textquotedblright~ \emph{Computing Research Repository (CoRR)},
Vol. abs/1310.1285, 2013.

\bibitem{39} http://www.nltk.org. 

\bibitem{40} A. Adhikari, S. Singh, B. Dutta, and A. Dutta,\textquotedblleft A novel information theoretic approach 
for finding semantic similarity in WordNet,\textquotedblright~\emph{In Proceeding of IEEE Region 10 Conference~TENCON}, Macau, China, pp. 1-6, 2015. 

\bibitem{41} P. Li, H. Wang, K. Q. Zhu, Z. Wang, X. Hu, and X. Wu,\textquotedblleft A Large Probabilistic Semantic Network Based
Approach to Compute Term Similarity,\textquotedblright~\emph{IEEE Transaction on Knowledge and Data Engineering}, VOL. 27, NO. 10, pp. 2604-2617, 2015.

\bibitem{42} G. A. Miller,\textquotedblleft WordNet: A lexical database for english,\textquotedblright~\emph{Communications of the ACM}, vol. 38, no. 11, pp. 39-41, 1995.

\bibitem{43} W. Wu, H. Li, H. Wang, and K. Q. Zhu,\textquotedblleft Probase: A probabilistic Taxonomy For Text Understanding,\textquotedblright~\emph{In 
Proceeding ACM International Conference on Management of Data (SIGMOD)}, Scottsdale, Arizona, USA,
pp. 481-492, 2012.

\bibitem{44} D. Bollegala, Y. Matsuo, and M. Ishizuka,\textquotedblleft A Web Search Engine-Based Approach to
Measure Semantic Similarity between Words,\textquotedblright~\emph{IEEE Transaction on Knowledge and Data Engineering}, VOL. 23, NO. 7, pp. 977-990, 2011.

\bibitem{45} A. Gledson and J. Keane,\textquotedblleft Using Web-Search Results to Measure
Word-Group Similarity,\textquotedblright~\emph{In Proceedings of the 22nd International Conference on Computational Linguistics, (COLING ’08)}, Manchester, 
United Kingdom,
Volume 1, pp. 281-288, 2008.

\bibitem{46} D. S\'{a}nchez, M. Batet, D. Isern, A. Valls, \textquotedblleft Ontology-based semantic similarity:
A new feature-based approach,\textquotedblright~\emph{
Expert Systems with Applications}, Vol. 39, pp. 7718-7728, 2012.

\bibitem{47} D. S\'{a}nchez, M. Batet,\textquotedblleft Semantic similarity estimation in the biomedical domain: An ontology-based
information-theoretic perspective,\textquotedblright ~\emph{Journal of Biomedical Informatics}, Vol. 44, pp. 749-759, 2011.

\bibitem{48} G. Pirr\'{o}, J. Euzenat,\textquotedblleft A Feature and Information Theoretic Framework
for Semantic Similarity and Relatedness,\textquotedblright~\emph{Chapter of book title The Semantic Web-ISWC 2010 of the series Lecture Notes in Computer Science}, 
Vol. 6496, pp. 615-630, 2010.

\bibitem{49} http://research.microsoft.com/en-us/projects/probase/.

\bibitem{50} Q. Yuan, Z. Yu, K. Wang, \textquotedblleft A New Model of Information Content for Measuring
the Semantic Similarity Between Concepts,\textquotedblright ~\emph{In Proceeding IEEE International Conference on Cloud Computing and Big Data}, pp. 141-146, 2013.

\bibitem{51} https://en.wikipedia.org/wiki/Discrete\_mathematics.

\bibitem{52} S. Harispe, D. S\'{a}nchez, S. Ranwez, S. Janaqi, and J. Montmain, \textquotedblleft A Framework For Unifying
Ontology-based Semantic Similarity Measures: A Study In The Biomedical Domain,\textquotedblright ~\emph{Journal of Biomedical Informatics}, Vol 48, pp. 38-53, 2014.

\bibitem{53} http://www.cs.technion.ac.il/\texttildelow gabr/resources/data/wordsim353
/wordsim353.html.

\bibitem{54} https://en.wikipedia.org/wiki/Semantic\_similarity.

\end{thebibliography}
\end{document}